\documentclass[pra,aps,twocolumn,showpacs]{revtex4-1}

\usepackage{graphicx}
\usepackage{bm,color}
\usepackage{physics}
\usepackage{amsmath, amssymb}
\usepackage{bm,color}
\usepackage{multirow}
\usepackage{ulem}

\newcommand{\be}{\begin{eqnarray}}
\newcommand{\ee}{\end{eqnarray}}

\begin{document}

\title{Information spreading and scrambling in disorder-free multiple-spin interacting models}
\date{\today}
\author{Yoshihito Kuno$^{1}$}
\author{Takahiro Orito$^{2}$}
\author{Ikuo Ichinose$^{3}$}

\affiliation{$^1$Graduate School of Engineering Science, Akita University, Akita 010-8502, Japan}
\affiliation{$^2$Graduate School of Advanced Science and Engineering, Hiroshima University, 739-8530, Japan}
\affiliation{$^3$Department of Applied Physics, Nagoya Institute of Technology, Nagoya, 466-8555, Japan}

\begin{abstract}
Tripartite mutual information (TMI) is an efficient observable to quantify the ability 
of scrambler for unitary time-evolution operator with quenched many-body Hamiltonian. 
In this paper, we give numerical demonstrations of the TMI in disorder-free
(translational invariant) spin models with 3-body and 4-body multiple-spin interactions. 
The dynamical behavior of the TMI of these models does {\it not} exhibit linear
light-cone for sufficiently strong interactions. 
In early-time evolution, the TMI displays distinct negative increase behavior fitted 
by a logarithmic-like function. 
This is in contrast to the conventional linear light-cone behavior present in 
the XXZ model and its near integrable vicinity. 
The late-time evolution of the TMI in finite-size systems is also numerically
investigated. 
The multiple-spin interactions make the system nearly integrable and weakly suppress 
the spread of information and scrambling.
The observation of the late-time value of the TMI
indicates that the scrambling nature of the system changes by interactions
and this change can be
characterized by a phase-transition-like behavior of the TMI, reflecting the system's
integrability and breaking of eigenstate thermalization hypothesis.
\end{abstract}


\maketitle
\section{Introduction}
Characterization of quantum chaos and scrambling (spread of information) \cite{Hayden,Sekino} is a topic of great interests
in broad physical fields from high-energy, condensed matter to information physics.
How information in a subsystem of an initial state spreads across the entire system through a quantum channel is currently studied extensively. 
In high energy physics, it was clarified that black hole has the most efficient ability of scrambling, spread of quantum information \cite{Sekino,Shenker2014,Maldacena2016}. 
In condensed matter physics, on the other hand,
how a many-body Hamiltonian, describing condensed matter phenomena, stirs quantum states under time evolution is a frequently asked question \cite{Nandkishore2015,Abanin}. 
In other words, which properties of the many-body Hamiltonian control the degree of 
spread of information is currently one of the most important topics of quantum physics. 
As typical examples, it is believed nowadays that Anderson localized system strictly inhibits the spread of information and scrambling, and certain many-body localized systems
also exhibit a similar nature~\cite{Basko2006,Nandkishore2015,Abanin}.
The above observation has been elucidated by studying the time evolution of bipartite entanglement entropy (BEE) \cite{Bardarson}. 
The notion of spread of information and scrambling is expected to be related to thermalization of the system~\cite{Rigol2007,Gogolin2016}, 
which is also an attractive topic in both high-energy and condensed matter physics. 
Intuitively, the realization of thermalization corresponds to the full spread of information and scrambling over the entire system under consideration, although there is no exact proof of it. 
In general, it is difficult to discuss universal relationships between
integrability, eigenstate thermalization hypothesis (ETH) and quantum information
propagation in a strict way. 
Therefore, it is important and desired to investigate these relationships in specific
models that can be analyzed by reliable methods.

So far, as an efficient tool to measure the spread of information and scrambling in
many-body systems, the out-of-time-ordered-correlation (OTOC) was proposed in \cite{Shenker2014,Maldacena2016}. 
The OTOC is an operator-based quantity to quantify the spread of information and
scrambling. 
The OTOC has been applied to various condensed matter systems, e.g., disordered system \cite{Swingle2017,Fan2017,Huang2017,Sahu2019} and some specific systems \cite{Bohrdt2017,Luitz2017,Lin2018,Knap2018,Hahn2021,Li2021,Colmenarez2020}.

Besides the OTOC, another quantity was proposed to quantify the degree of scrambling,
which extracts the scrambling ability of the operator of the quantum channel itself: 
The tripartite mutual information (TMI) of a dual pure state obtained by the 
state-channel map, proposed by Hosur, et. al.~\cite{Hosur}. 
The negativity of the TMI indicates the non-locality of information, i.e., 
the spread of information and scrambling across the entire system. The TMI is an efficient quantitative diagnostic of the spread of information and 
scrambling especially for unitary time-evolution operator $U(t)=e^{-itH}$,
where $H$ is a many-body Hamiltonian. 
The spread of information and scrambling may be well observed not as the dynamics of the many-body wavefunction, but as one of the properties of time-evolution operator of a many-body Hamiltonian \cite{Zhou2017}. 
We note that a choice of observable is needed for calculating the OTOC, 
but for the TMI, it is not. 
While we often take ensemble of random initial product states in the conventional measure of the BEE, the calculation of the TMI does not require that procedure.

The TMI in Ref.~\cite{Hosur}
is a physical quantity that reflects the unitary time-evolution operator directly, 
by treating all states on an equal footing as initial states,
although it is computationally expensive. 
However, the study of the systematic observation of the TMI for various many-body systems
is still lacking. 
It is therefore important to investigate 
how scrambling changes with the different physical properties of individual models 
by observing the TMI and to clarify relevant ingredients for scrambling. 

We shall study the TMI for disorder-free spin models with 
multiple-spin interactions, each of which is an interesting model exhibiting
nonlocalized integrable, non-integrable and local integrable properties depending
on the strength of the interactions. 
In this paper, two types of model are investigated, which are generalizations of the standard 
$s=\frac{1}{2}$ XXZ spin model:
(I) Spin model with 3-body interactions;
(II) Model including 4-range multiple-spin interactions.  
In the previous study \cite{Michailidis2018}, the BEE was studied for similar models 
to the above by employing a specific type of initial state,
and the emergence of slow thermalization was observed there.
However, how the quench time-evolution operator 
influences the spread of information and scrambling has not been observed. 

Although a recent study \cite{MacCormack2021} gave a general classification for the late-time dynamics of the TMI in various types of model, recent numerical calculations of the late time value (saturation value) of the TMI in certain models \cite{Iyoda2018,Schnaack2019,Wanisch2021} indicate
no correlations between integrability of the system and dynamics of the TMI. 
For example in the Ref.\cite{Iyoda2018}, even for integrable system the
scrambling characteristics obtained by the TMI depend distinctly on the choice of initial
state, Neel or all up-spin states. 
Also in Ref.~\cite{Schnaack2019}, the spinless
interacting fermion, which is an integrable model, exhibits strong scrambling for
strong interactions.
On the other hand, other studies of the TMI for the 
many-body localized (MBL) systems \cite{Bolter2021,Mascot2020} imply the existence of
certain relationship between them. 
Actually, this discrepancy comes from the difference in the settings on calculating 
the TMI in various works, such as employing specific initial states, state-partitioning,
etc.
Among them, the TMI defined in the doubled-Hilbert space through the state-channel
map can be efficient tool free from the choice of specific initial states. 

As we explain in Sec.~II, the doubled Hilbert-space TMI describes the scrambling in
a reliable manner, and therefore,
it is important to investigate relevant models using the TMI introduced in
Ref.~\cite{Hosur} to get reliable insight into the above-mentioned relationship. 
In fact,
studies of scrambling properties of various relevant models can answer the very
question if there exists universal and general relationship between
quantum spread of information, near-integrability, ETH and localization, as
we cannot exclude the possibility that there are no such universal relations between them,
and each model exhibits its own properties of scrambling.
This is obviously an open question to be studied by the TMI, 
which is addressed in this work.

Motivated by the above observations, we shall investigate the behavior of the TMI 
for the specific 3-body and 4-range models.
In particular, as the target models reduce to the integrable XXZ model for
the vanishing multiple-spin interactions and also they acquire integrability 
for the strong-coupling limit, the present study on the models reveals some
properties of the phases in the vicinity of integrability, and it may also give useful
insight into finite-size (intermediate) MBL regime and chaos due to breaking
integrability, which were proposed recently \cite{morningstar,Bulchandani}. 
Besides these works, there appeared several interesting studies on weakly
broken integrability phenomena these 
days~\cite{Brenes,Znidaric,Friedman, Durnin,Hutsalyuk},
whose relationship with the present work is an interesting future problem.

In this paper, by the numerical study of the TMI, 
we obtain the following two observations: 
(I) For the standard XXZ model, the TMI in early-time evolution exhibits linear light-cone spread. 
On the other hand for the models with multiple-spin interactions, we find that the decrease of the TMI (an increase of the absolute value of the TMI) in early-time evolution is fitted in a logarithmic-like function of time, that is, the multiple-spin interactions change the behavior of the TMI. 
And also, the spatial profile of the TMI reflects these observations. 
The change of the spreading nature is related to the properties of the integrability and
ETH in the models obtained by the level spacing analysis (LSA) and observing 
the histograms of local magnetization. 
(II) We investigate the late-time evolution of the TMI. 
The present finite-size system calculation indicates that 
the scrambling nature characterized by the late time saturation value of the TMI
correlates well to the properties of integrability and ETH of the models, 
which are revealed by the LSA and the histograms of local magnetization.

By observing the very late-time value of the TMI, 
we find that the near saturation values exhibit phase-transition-like behavior
where the scrambling nature changes by increasing the strength of the multi-spin
interactions. 
This transition behavior of the TMI is also related to the properties of 
integrability and ETH of the models, which are revealed by the LSA and the histograms 
of local magnetization.

The rest of this paper is organized as follows.
In Sec.~II, we introduce the TMI and explain the methods for the practical calculation. 
In Sec.~III, we introduce the target disorder-free spin models and also
briefly explain their physical properties, which have been revealed by the previous work. 
In Sec.~IV, we investigate the eigenvalue and eigenstate properties of the models
to elucidate integrability and the character of ETH.
These results give insight into the nature of spreading of the quantum information 
by observing the TMI. 
Then in Sec.V and Sec.VI, we show the numerical calculations of the TMI for the target spin models. 
In particular, the numerical results for two different time intervals are shown,
i.e., early-time evolution and late-time ones, and further show the spatial profile of the TMI in a specific setup. 
We discuss the physical meanings of the results.
In Sec.~VII, we investigate the nearly saturation values of the TMI and show usefulness 
of the TMI to characterize a phase transition, which is consistent to both integrability
and ETH properties estimated in Sec.~IV. 
Section VIII is devoted to discussion and conclusion.


\section{Calculation of TMI}
\begin{figure}[t]
\begin{center} 
\includegraphics[width=7cm]{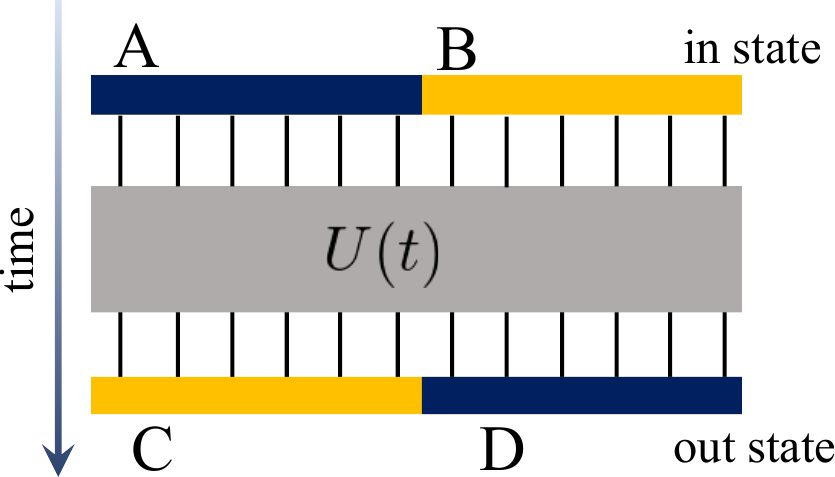}  
\end{center} 
\caption{Schematic image of the time evolution of the state with doubled Hilbert space. The spatial partitioning of the system is represented where four subsystems A, B, C, and D are introduced. Each part is $L/2$-lattice sites.}
\label{Fig1}
\end{figure}
In this section, we introduce the TMI proposed in \cite{Hosur}, and briefly explain the methods of the practical numerical calculation to be applied for
one-dimensional lattice models with $L$ sites. 
Our numerical resource allows us to calculate the TMI up to the system size $L=14$ by the methods.

We investigate properties of the spread of information and scrambling embedded 
in the time evolution operator $\hat{U}(t)\equiv e^{-itH}$. 
In the treatment of the time-evolution operator in calculating the TMI,
the state-channel map plays an essential role. 
Under this map, the time evolution operator $\hat{U}(t)\equiv e^{-itH}$ 
is regarded as a pure state in the doubled Hilbert space,
${\cal H}_{\rm D} \equiv{\cal H}_{\rm in} \otimes {\cal H}_{\rm out}$~\cite{Hosur}.
That is, we start from the density matrix at time $t$, $\rho(t)=\sum^{N_D}_{\nu=1}p_{\nu}\hat{U}(t)|\nu\rangle \langle\nu|({\hat{U}(t)})^{\dagger}$, where $\{|\nu\rangle\}$ is 
a set of a orthogonal bases state (time independent),
$N_D$ is the dimension of the Hilbert space in the system, and an arbitrary 
input ensemble is tuned by parameters $\{ p_\nu\}$. 
Then, by the state-channel map applied to this density matrix $\rho(t)$,
the operator can be mapped into a pure state in the doubled Hilbert space,
\begin{eqnarray}
\rho(t)\to 
|U(t)\rangle=\sum_{\mu}\sqrt{p_\nu}(\hat{I}\otimes {\hat U}(t))
|\nu\rangle_{\rm in}\otimes|\nu\rangle_{\rm out},
\label{pure_state_U}
\end{eqnarray}
where $\hat{I}$ is the identity operator and $\{|\nu\rangle_{\rm in}\}$ and
$\{|\nu\rangle_{\rm out}\}$ are the same set of orthogonal bases state, and therefore,
the state is defined on the doubled Hilbert space, ${\cal H}_{\rm D}$, spanned by 
$\{|\nu\rangle_{\rm in}\}\otimes \{|\nu\rangle_{\rm out}\}$. 
The time evolution operator ${\hat U}(t)$ acts only on the out orthogonal states $|\nu\rangle_{\rm out}$.
Even though arbitrary input ensemble can be employed by tuning $\{ p_\nu\}$ \cite{Hosur}, 
in this work, we focus on the infinite temperature case, such as ${p_{\nu}}=1/N_{D}$. 
Then, for initial state at $t=0$, $\hat{U}(0)=\hat{I}$, then the in-state and 
out-state  are maximally entangled. 

To estimate the TMI for the spread of information and scrambling in the time evolution, 
we introduce spatial partitioning to the pure state $|U(t)\rangle$. 
The spatial partitioning is done for both the $t=0$ in-state and the out-state at $t$. 
As shown in Fig.~\ref{Fig1}, the $t=0$ state (given by $\rho(t=0)$) is divided into two subsystems $A$ and $B$, 
and similarly the state at time $t$ (given by $\rho(t)$) is divided into two subsystems $C$ and $D$.
We mostly employ the partition with the equal length of $A$ and $B$ ($C$ and $D$)
subsystems for the practical calculation,
although some specific partitioning will be used for studying 
spatial pattern of scrambling in Sec.~VI.

Under this spatial partitioning, the density matrix of the pure state 
$|U(t)\rangle \in {\cal H}_{\rm D}$ 
is defined as $\rho_{ABCD}(t)=|U(t)\rangle \langle U(t)|$.
From this full density matrix $\rho_{ABCD}(t)$, 
a reduced density matrix for a subsystem $X$ is obtained by tracing out the degrees of freedom in the complementary subsystem of $X$ denoted by ${\bar X}$, i.e., $\rho_{X}(t)=\mathrm{tr}_{\bar X}\rho_{ABCD}$. 
From the reduced density matrix $\rho_{X}(t)$, the operator entanglement entropy (OEE) for 
the subsystem $X$ is obtained by conventional von-Neumann entanglement entropy, 
$S_X=-\mathrm{tr}[\rho_{X}\log_2 \rho_{X}]$.

From the OEE, we define the mutual information between $X$ and $Y$ subsystems 
(where $X, Y$ are some elements of the set of the subsystems $\{A,B,C,D\}$, 
and $X\neq Y$); 
\begin{eqnarray}
I(X:Y)=S_X+S_Y-S_{XY}.
\label{MI}
\end{eqnarray}
This quantity indicates how the subsystems $X$ and $Y$ correlate with each other.

By using the mutual information, the TMI for the subsystems $A$, $C$ and $D$ is 
defined as; 
\begin{eqnarray}
I_3(A:C:D)=I(A:C)+I(A:D)-I(A:CD).
\label{TMI_def}
\end{eqnarray}
This quantity is a measure for how the initial information embedded in 
the subsystem $A$ spreads into both subsystems $C$ and $D$ in the output state. 
If the spread of the information in $A$ sufficiently occurs across the entire system
at time $t$, $I_3(t)$ gets negative, while 
the mutual information keeps a non-negative value even in such a situation.
Then, as proposed in Ref.~\cite{Hosur}, the TMI, $I_3$, can be used to quantify
the degree of scrambling, i.e., 
the spread of information is characterized by a negative value of $I_3$.
In general $I_3$ is zero at $t=0$, as $|U(0)\rangle$ is the product state of the
EPR pair at each lattice site.
Then, if the time-evolution operator acts as a strong scrambler, $I_3$ acquires a large
negative value under the time evolution. 
On the other hand, if the time evolution operator
does not act as a scrambler, $I_3$ remains small.  
Hence, $I_3$ is a good indicator to quantify the degree of scrambling, i.e., the spread of information. 
In this paper, we mostly employ the TMI to characterize the scrambling for our target models.

Here, we explain the practical numerical calculation of the TMI. 
The numerical cost for the straightforward manipulation of the density matrix
$\hat{\rho}_{ABCD}$ is quite high. 
Instead, we make use of the singular value decomposition (SVD) to the pure state
$|U(t)\rangle$. 
For a certain partitioning $X$ and ${\bar X}$, the pure state is written as
\begin{eqnarray}
|U(t)\rangle&=&\frac{1}{N_D}\sum_{\nu}(\hat{I}\otimes {\hat U}(t))|\nu\rangle_{\rm in}|\nu\rangle_{\rm out}\nonumber\\
&=&\frac{1}{N_D}\sum_{k_X,\ell_{\bar X}}U_{k_X,\ell_{\bar X}}
|k_{X}\rangle_{X} |\ell_{\bar X}\rangle_{\bar X}\nonumber\\
&\stackrel{SVD}{=}&\sum_{r}\lambda^{X,\bar{X}}_{r}|r\rangle_{X} |r\rangle_{\bar X}.
\label{TMI_cal1}
\end{eqnarray}
Here, in the second line, the input and output basis states are reassembled into basis
vectors $\{|k_{X}\rangle_X\}$ and $\{|\ell_{\bar X}\rangle_{\bar X}\}$ corresponding to the spatial partition $X$ and $\bar{X}$, and then, 
a concrete matrix representation of the operator $(\hat{I}\otimes {\hat U}(t))$ is obtained.
In the third line, we simply carry out the SVD to obtain singular values,
$\lambda^{X,\bar{X}}_{r}$,
and the OEE for the subsystem $X$ is straightforwardly obtained by
$S_{X}=-\sum_{r}(\lambda^{X,\bar{X}}_{r})^2\log_2 (\lambda^{X,\bar{X}}_{r})^2$.
Hence, from the numerical calculation of OEE, we evaluate the TMI, $I_3$.

In the following numerical calculations, 
we focus on spatially equal-partitioning case: i.e.,
as we briefly mentioned in the above,
the subsystems $A$ and $B$ are defined as the $L/2$-site left and right subsystems in 
the input state, respectively, and the subsystems $C$ and $D$ are defined similarly
as the $L/2$-site systems in the out-put state as shown in Fig.~\ref{Fig1}. 
We also focus on the zero magnetization sector of the Hilbert space in the choice of
the set of bases $\{|\nu\rangle_{\rm in(out)}\}$. 
Under this setup, $S_{Y}$ with $Y=A,B,C$ and $D$ is a constant at any time, as their values are shown in Appendix A.
We further set a reference frame of the TMI, $I_3$, as in Refs.~\cite{Schnaack2019,Bolter2021}. 
The reference flame is the value of the TMI of the Haar random unitary, $I^{H}_{3}$, 
which depends on the Hilbert space dimension of the system size $L$ \cite{Haar_ND}. 
The value of $I^{H}_{3}$ can be numerically calculated. 
Then, we define a normalized TMI, $\tilde{I}_3(A:C:D)$ as follows,
\begin{eqnarray}
\tilde{I}_3(A:C:D)\equiv \frac{I_3(t)-I_{3}(0)}{|I^{H}_{3}-I_{3}(0)|}.
\label{TMI_def}
\end{eqnarray}
In the following sections, we numerically obtain the value of $\tilde{I}_3$.

We here comment on the saturation of the TMI in a strong scrambling case. 
As explained in Ref.~\cite{Hosur}, even for strong scrambling, 
the TMI of the time-evolution operator without fixing magnetization sector 
does not reach the Haar-scrambled limit, $\tilde{I}_3=1$. 
Furthermore, since we focus on the zero-magnetization sector,
the value of the TMI tends to be lowered by the constraint of the sector, 
however, nonetheless, the saturation value of the TMI exhibits the characteristic 
behavior depending on the model parameters as we show in the following.

As another comment, we would like to note our recent work on quantum spin models
with topological order~\cite{Orito2022}.
There, we found that results obtained by calculating the TMI are quite stable and 
reliable compared with those by the quench EE.
Therefore, we can regard the TMI as a benchmark for observation of the scrambling.

\section{Target models}

In this paper, we consider three spin models: the XXZ model, 3-body spin model, 
and 4-range model. 
These models are given as follows; 
\begin{eqnarray}
&&H_{\rm XXZ}=\sum_{j}J_{1}S^{z}_jS^{z}_{j+1}+H_{\rm hop},\label{H_XXZ}\\
&&H_{\rm 3B}=\sum_{j}J_{1}S^{z}_jS^{z}_{j+1}+J_2S^{z}_{j}S^{z}_{j+2}
+J_3S^{z}_{j}S^{z}_{j+1}S^{z}_{j+2}\nonumber\\ 
&&\hspace{1.5cm} +H_{\rm hop},\\
&&H_{\rm 4R}=\sum^{4}_{\alpha=2}t_{\alpha}\hat{h}_{\alpha}+H_{\rm hop},
\label{3B_model}
\end{eqnarray}
where 
\begin{eqnarray}
H_{\rm hop}=\frac{v}{2}\sum_{j}(S^{+}_jS^{-}_{j+1}+S^{-}_jS^{+}_{j+1}),\nonumber
\end{eqnarray}
and $J_i\; (i=1,2,3)$ and also $v$ in $H_{\rm hop}$ are parameters.
The parameter $\alpha$ in the model $H_{\rm 4R}$ [in Eq.~(\ref{3B_model})]
denotes the range of the interactions and each $\hat{h}_{\alpha}$ is given in TABLE I.
The XXZ model, $H_{\rm XXZ}$, is a nonlocalized integrable model, and 
only a tiny integrability-breaking perturbation makes the model satisfy ETH~\cite{Brenes}.

\begin{table}[t]
\centering
\begin{tabular}{ |c||c|c| }
\hline
${\hat h}_\alpha$ & {Included terms}\\
\hline
2-body, $\hat{h}_2$ & $S^{z}_{j}S^{z}_{j+1},\: S^{z}_{j}S^{z}_{j+2},\:S^{z}_{j}S^{z}_{j+3}$\\
\hline
3-body, $\hat{h}_3$ & $S^{z}_{j}S^{z}_{j+1}S^{z}_{j+2},\: S^{z}_{j}S^{z}_{j+2}S^{z}_{j+3},\:S^{z}_{j}S^{z}_{j+1}S^{z}_{j+3}$\\
\hline
4-body, $\hat{h}_4$ & $S^{z}_{j}S^{z}_{j+1}S^{z}_{j+2}S^{z}_{j+3}$ \\ 
\hline
\end{tabular}
\caption{Included terms for each $\alpha$-body Hamiltonian, ${\hat h}_{\alpha}$.}
\label{table:1}
\end{table}

Experimentally, the 3-body model can be feasible in an effective theory of the
Bose-Hubbard model describing cold atoms on a zig-zag optical lattice \cite{Pachos2004},
where the 3-body terms perturbatively appear, and also the 3-body terms 
can be implemented experimentally in cold polar molecules \cite{Buchler2007}.

For any value of $v$, the XXZ model $H_{\rm XXZ}$ is integrable. 
For $v=0$, the remaining two models, $H_{\rm 3B}$ and $H_{\rm 4R}$, are integrable 
since each eigenstate is characterized by on-site (local) conserved quantities,
i.e., the eigenvalues $\pm 1/2$ of $\{S^{z}_j\}$ 
coming from the fact $[H_{{\rm 3B}({\rm 4R})},S^{z}_j]=0$ for any $j$. 
In addition, there exist multiple-site local conserved quantities such as $S^{z}_{j}S^{z}_{j+1}S^{z}_{j+2}$ and $S^{z}_{j}S^{z}_{j+1}S^{z}_{j+2}S^{z}_{j+3}$, etc. 
These can be regarded as short domain wall operators. 
Such term may induce a quasi-localization phenomenon, namely, Hilbert space fragmentation or shattering \cite{DeTomasi2019,Santos2021,Sala2020}. 
The existence of the `hopping' $H_{\rm hop}$ breaks the integrability of the two models
$H_{\rm 3B}$ and $H_{\rm 4R}$, 
that is, these models with a tiny but finite $v$ turn to non-integrable in a strict sense.
Sometimes they are called nearly integrable model. 

Previous study \cite{Michailidis2018} showed that the 3-body and 4-range models 
display slow-thermalization for sufficiently small $v$. 
In particular, it was numerically demonstrated that the 3-body model exhibits 
a slow increase of the BEE for initial product states. 
The presence of the $J_2$ and $J_3$-terms {\it hinders} 
the growth of the entanglement entropy.
Also, interestingly enough, the early-time evolution of the entanglement entropy 
displays a logarithmic-like curve in time, which 
is in contrast to the standard linear-light cone increase in the standard XXZ model.

It is important to investigate how such a slowing-down or unconventional behavior
of the time evolution of the systems, induced by multiple-spin interactions, 
reflects quantum information scrambling measured by the TMI. 
Furthermore, one may wonder how the magnitude of the hopping term, $H_{\rm hop}$, 
changes the dynamics of the TMI for both 3-body and 4-range models.  
We shall address these problems by the numerical methods in the following sections. 

In what follows, we put $\hbar=1$ and set the parameters as
$v=0.2$ and for $H_{\rm XXZ}$ and $H_{\rm 3B}$, $J_1=0.3$, and 
for $H_{\rm 3B}$, $J_2=J_3\equiv J_0$ with a varying 
parameter $J_0$. For $H_{\rm 4R}$, we set all the couplings to the same value, such as $t_2=t_3=t_4\equiv t_0$.

\section{Integrable and localization property}
Before going to the numerical calculations of the TMI, 
we study the integrability and localization characters of the models by the level spacing
analysis (LSA) \cite{Oganesyan2007,Pal2010} 
and observing the histograms of the local magnetizations \cite{Laflorencie2020}. 
These results give an insight into the spreading behavior of the quantum information,
which is examined using the TMI in later sections.

\subsection{Level spacing analysis}
\begin{figure}[t]
\begin{center} 
\includegraphics[width=8.5cm]{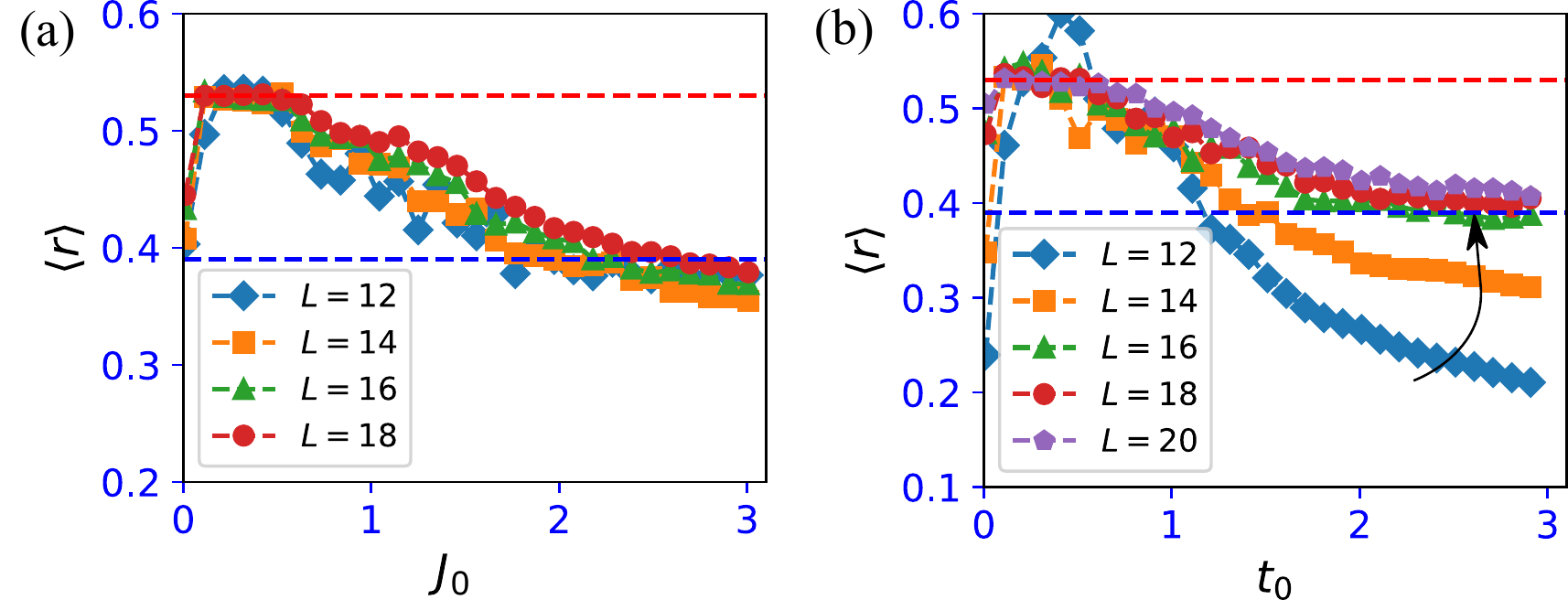}  
\end{center} 
\caption{Level spacing ratio for 3-body model[(a)] and 4-range model [(b)]. 
We set $L=12$-$18$ [(a)] and $L=12$-$20$ [(b)], $J_1=0.3$ and $v=0.2$. 
The red and blue dashed lines represent the WD value $0.53$ and the Poisson value $0.39$.}
\label{LSA}
\end{figure}

We investigate the near integrability of the 3-body and 4-range models by using the LSA. 
Whether the models are 
nearly integrable or not is determined by the values of $J_2$, and $J_3$ for the 3-body
model and the values of $t_{\alpha}$ for the 4-range model. 
Since we expect that the integrability properties are not influenced by the boundary 
condition, we employ periodic boundary condition (PBC).

To extract the integrable properties of the model straightforwardly, 
we diagonalize the Hamiltonian of the models in a sector with a fixed momentum and 
the positive parity, since the models are invariant under spatial translation and
also invariant under the parity operation.
Here we employ Quspin solver~\cite{Quspin} to diagonalize the Hamiltonian.
Then we take all eigenvalues of Hamiltonian in each momentum sector 
and calculate the level spacing
$r^{k}_s$ defined by $r^k_{s}
=[{\rm min}(\delta^{(s)}_k, \delta^{(s+1)})_k]/[{\rm max}(\delta^{(s)_k},\delta^{(s+1)})]$
for $s$, where $\delta^{(s)}_k=E^k_{s+1}-E^k_{s}$ 
and 
$\{E^k_{s}\}$ is the set of energy eigenvalues in ascending order in momentum sector $k$ 
and $s$ labels the elements of eigenvalues of the Hamiltonian in momentum sector 
$k$. 
We average over the suffix $s$ and obtain each mean-level spacing $\langle r\rangle_{k}$ 
in each momentum sector.

In general, if the system is integrable, 
the average level spacing takes $\langle r\rangle\simeq 0.39$, corresponding to 
the Poisson distribution. 
On the other hand, if the system is non-integrable (chaotic), the average level 
spacing takes
$\langle r\rangle\simeq 0.53$, corresponding to the Wigner-Dyson (WD) distribution \cite{Oganesyan2007,Pal2010}.

By varying $J_0$ for the 3-body model and $t_0$ for the 4-range model,
we observe how $\langle r\rangle$ behaves. 
Figure \ref{LSA} (a) shows the result of the 3-body case. 
Here, we average over the suffix $k$ 
and obtain the total averaged value over $k$, $\langle r\rangle$. 
For small $J_0$, where the hopping term is dominant, 
$\langle r\rangle$ is close to the value of the WD distribution. 
Thus, the system is non-integrable. 
As increasing $J_0$, we observe that $\langle r\rangle$ approaches the value of the Poisson distribution. This indicates that the 3-body model approaches being integrable.

Next, we turn to the results of the 4-range model in Fig.~\ref{LSA} (b). 
Here, we show the result of zero momentum sector $k=0$. 
For small $t_{0}$, $\langle r\rangle$ is close to the value of the WD distribution. 
But the deviation is large for small $L$. 
Entirely, the system tends to be non-integrable. 
As increasing $t_{0}$, we observe that $\langle r\rangle$ deviates from the value of 
the WD distribution. 
For large $L$ the value well approaches the value of the Poisson distribution. 
For small $L$, $\langle r\rangle$ decreases up to a value less than that of the Poisson
distribution.
Hence, in the thermodynamic limit, the system is integrable for large $t_0$. 
Even for small $L$ and large $t_0$, the system is not chaotic at least.

The above LSA results are fairly in good agreement with the results in the previous 
work \cite{Michailidis2018}, and will give a useful insight into
the numerical results of the TMI, which are obtained later on.
\begin{figure}[t]
\begin{center} 
\includegraphics[width=8.5cm]{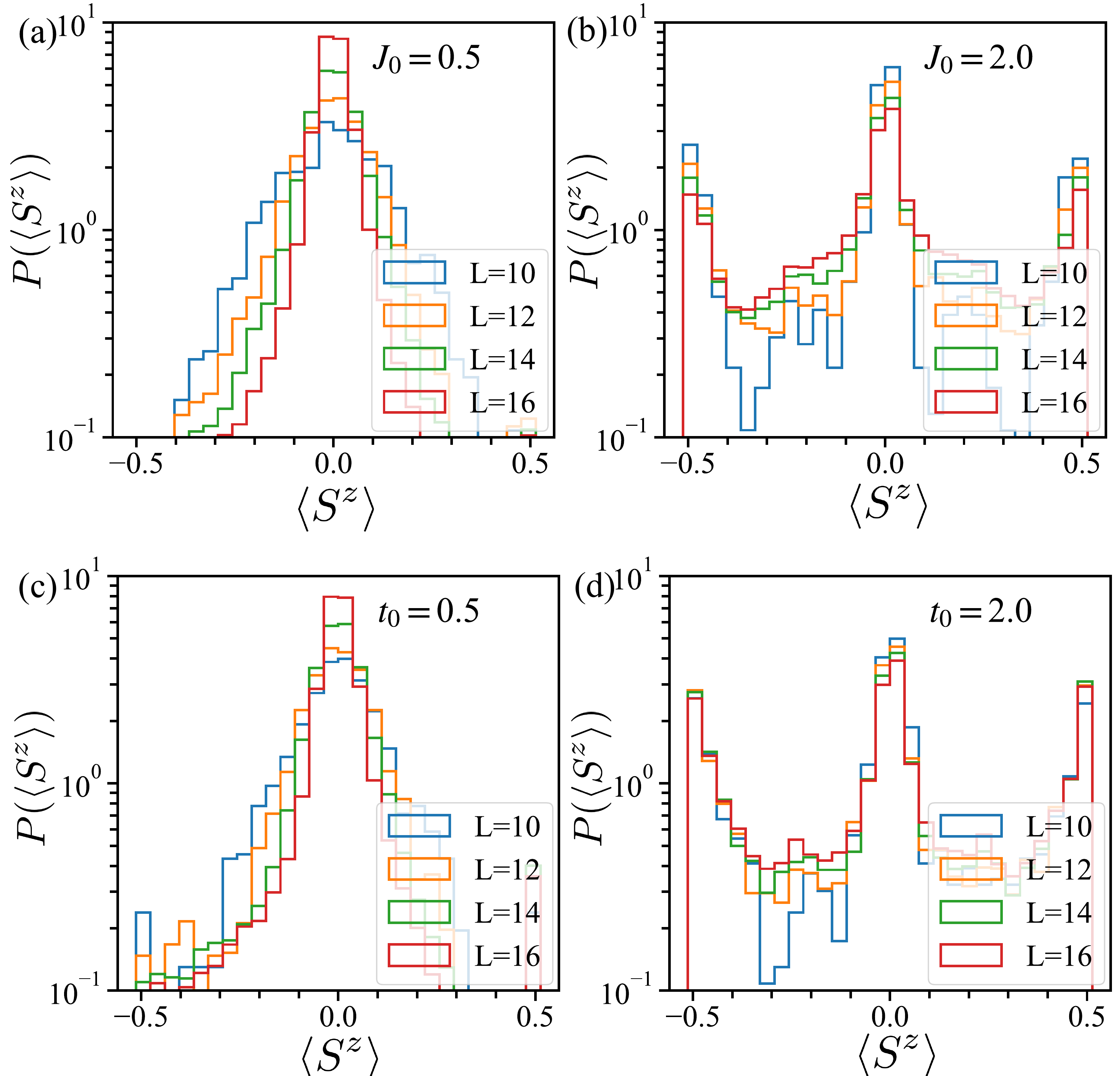}  
\end{center} 
\caption{Histograms of the local magnetization in the 3-body and 4-range models. We employ all eigenstates for $L=10$-$16$.
(a) $J_0=0.5$, 3-body model case, where the model is non-integrable as for the LSA. (b) $J_0=2$, 3-body model case.
(c) $t_0=0.5$, 4-range model case, where the model is non-integrable as for the LSA. (d) $J_0=2$, 4-range model case.}
\label{H_LSz}
\end{figure}

\subsection{Local magnetization}
To obtain an intuitive picture of the models, 
we furthermore measure the local magnetization of every eigenstates of the 3-body and 
4-range models. 
The histograms of the local magnetization $\langle S^{z}\rangle$, obtained from
the wave function of eigenstates in the 3-body and 4-range models, give useful 
information on localization tendency \cite{Laflorencie2020,Dupont2019,Hopjan2020}, i.e., 
from this observation, we can see whether the system satisfies the ETH or not. 
If it has a Gaussian distribution with its peak at $\langle S^{z}\rangle=0$, it is 
expected that the model satisfies the ETH. 
On the other hand, if it has two peaks at 
$\langle S^{z}\rangle=\pm 1/2$, the ETH breaks down there. 
Hence from the histograms, we can judge whether the system has a localization nature
or not.
Furthermore, the shape of the histograms reveals the degree of the spreading of quantum
information in the models.

Figure \ref{H_LSz} displays the histograms $P(\langle S^{z}\rangle)$ for $J_0=0.5$ [Fig.~\ref{H_LSz}(a)] and $J_0=2$ [Fig.~\ref{H_LSz}(b)] in the 3-body model, 
and for $t_0=0.5$ [Fig.~\ref{H_LSz}(c)] and $t_0=2$ [Fig.~\ref{H_LSz}(d)] in 
the 4-body model. 
From $J_0=0.5$ and $t_0=0.5$ data of the 3-body and 4-range models, the ETH seems to 
be satisfied as the $P(\langle S^{z}\rangle)$ has the Gaussian-like distribution 
with its peak at $\langle S^{z}\rangle=0$. 
For $J_0=2$ and $t_0=2$ cases, on the other hand, the distribution is not a standard one,
that is, there are three peaks at $\langle S^{z}\rangle=0$ and $\pm 1/2$. 
This result implies that the strong ETH breaks down and the systems have mixed
properties of the ETH and non-ETH. 
Even in a change of the system size $L$, the heights of the three peaks are almost
the same. 
From this result, we expect for large $J_0$ and $t_0$ regimes, weak localization
takes place there, which gives some effects on the behavior of the TMI.

The above non-standard behavior of the local magnetization can be
understood in the following way.
For the 3-body model with large $J_0\gg J_1, v$, the terms $\{S^z_jS^z_{j+2}\}$
and also $\{S^z_jS^z_{j+1}S^z_{j+2}\}$ dominate and they act similarly to LIOMs 
in the MBL regime.
In energy eigenstates, some of 2-body terms acquire $S^z_jS^z_{j+2} \sim \pm 1/4$, and
then, the 3-body term generates a `random potential' such as
$S^z_jS^z_{j+1}S^z_{j+2}\sim \pm 1/4 S^z_{j+1}$.
As a result, $S^z_{j+1} \sim \pm 1/2$.
On the other hand, the peak at $\langle S^z\rangle \sim 0$ emerges as a result
of the frustration between the 2-body and 3-body terms, i.e., 
$\langle S^z_jS^z_{j+2}\rangle \sim 0$ and/or 
$\langle S^z_jS^z_{j+1}S^z_{j+2}\rangle \sim 0$ by the linear
combination of $S^z_j$ eigenstates.\\

\begin{figure}[t]
\begin{center} 
\includegraphics[width=8.5cm]{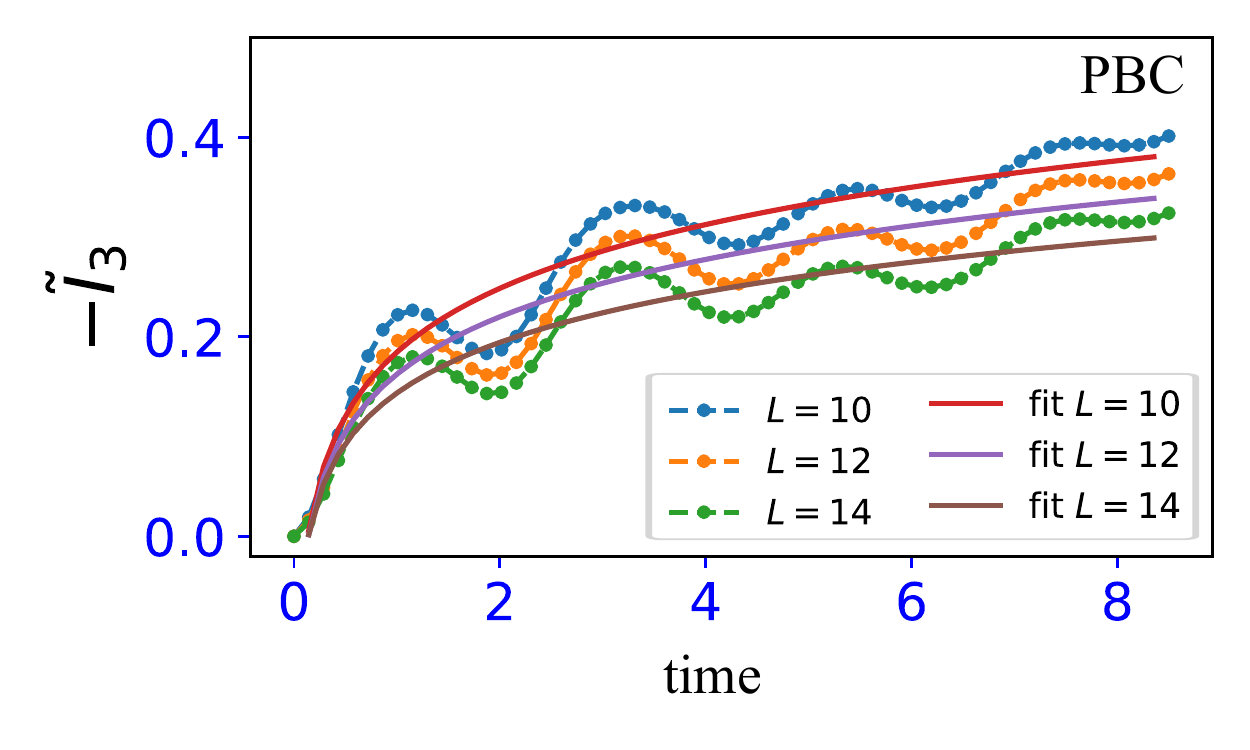}  
\end{center} 
\caption{Early-time evolution of 3-body model $H_{\rm 3B}$. 
The logarithmic fitting lines are 
$(-\tilde{I}_3)=0.0639 \log_2 t+0.184$, $(-\tilde{I}_3)=0.0575 \log_2 t+0.1624$ and
$(-\tilde{I}_3)=0.0507 \log_2 t +0.1435$ for $L=10$, $12$ and $14$, respectively.
We set $J_1=0.3$, $J_0=2$ and $v=0.2$.
In the fitting, we used the data points within $t\in[0.1:8]$.
The unit of time is $[2\hbar/v]$.}
\label{Fig2}
\end{figure}


\section{Numerical investigation of dynamics of TMI}

In this section, we shall show numerical results of the TMI for the models,
$H_{\rm XXZ}$, $H_{\rm 3B}$ and $H_{\rm 4R}$. We employ Quspin package \cite{Quspin} 
in generating the time evolution operator of the target Hamiltonians.
We focus on two observing time intervals: (I) early-time evolution where the time interval is set $t_I \leq 10$, 
which is less than the time $(\hbar L)/2({\rm max}(v/2,J_1,J_0))$ at
which an excitation from the center of the system almost reaches the edges. 
(II) late-time evolution where the time interval is set $t_{II} \leq 200$, much larger than the time $(\hbar L)/2({\rm max}(v/2,J_1,J_0))$. 
During that time interval, the TMI almost saturates as shown later.
In our numerical resource, the accessible system size of the calculation of the TMI 
is up to $L=14$ and both open and periodic boundary conditions (OBC and PBC) are employed.
We think that the early-time evolution is not affected substantially by the finite-size
and boundary effects. 
On the other hand, the long-time evolution is affected by both of them, 
but we expect that the data give useful insight into the spread of information in 
a finite-size system since some experimental systems to be prepared to simulate our 
target spin models is obviously a finite-size system.

\subsection{Early-time evolution of TMI in XXZ and 3-body models}

Let us move on numerical calculation of the TMI in early-time evolution for the XXZ 
and 3-body models. 
For the 3-body model of $H_{3B}$, we set $J_0=2$, 
where the LSA of the 3-body model indicates that the model 
is in the nearly-integrable regime (See Fig.\ref{LSA} (a)).
It is interesting how such near-integrability reflects the early-time evolution 
of the TMI. 
As a reference value of the time evolution of the TMI, we use values of 
the Haar random unitary numerically obtained as,
$I^{H}_{3}$, $I^{H}_{3}/L=-8.559$, $-10.5576$ and $-12.5566$ for $L=10$, $12$ and $14$,
respectively.

\begin{figure}[t]
\begin{center} 
\includegraphics[width=8.5cm]{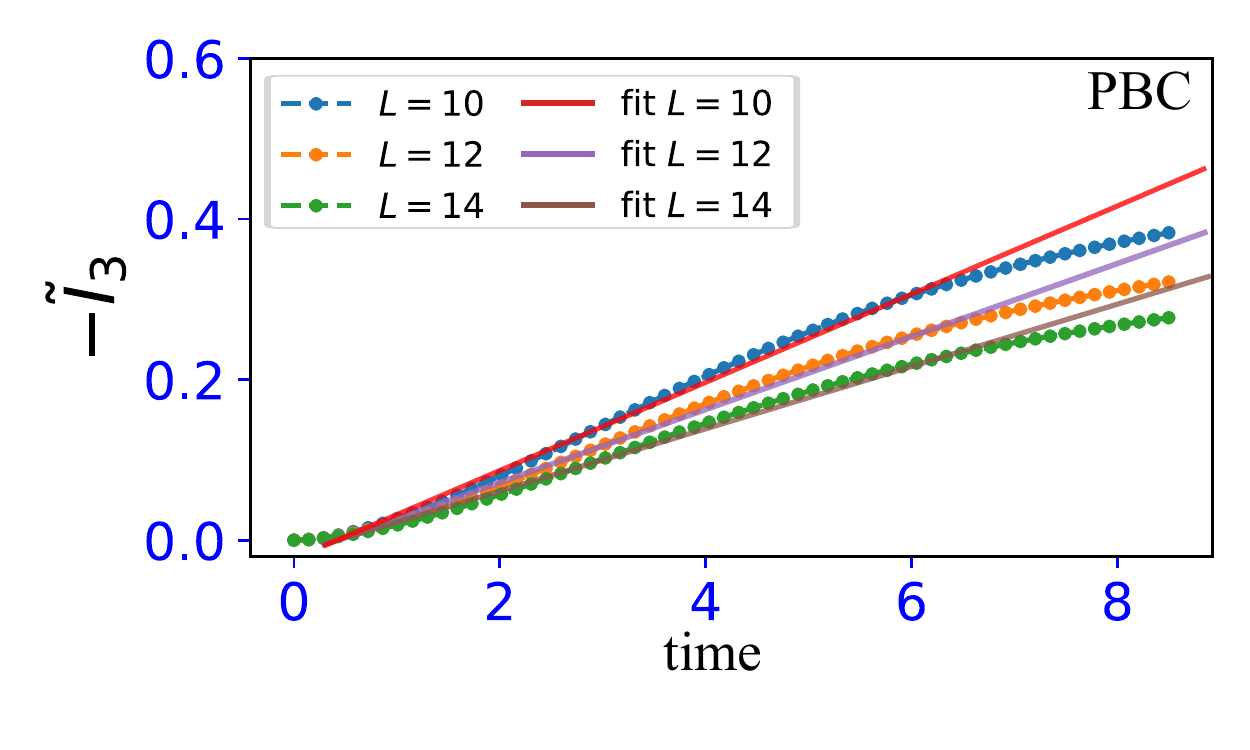}  
\end{center} 
\caption{Early-time evolution of XXZ model. 
The linear fitting lines are $(-\tilde{I}_3)=0.0549 t-0.0231$, 
$(-\tilde{I}_3)=0.0459 t-0.0199$ and
$(-\tilde{I}_3)=0.0395 t -0.0174$ for $L=10$, $12$ and $14$, respectively. 
We set $v=0.2$ and $J_1=0.3$.
In the fitting, we used the data points within $t\in[0.28:6.94]$. The unit of time is $[2\hbar/v]$.}
\label{Fig3}
\end{figure}

Figure \ref{Fig2} shows the early-time evolution of $\tilde{I}_3$ in the 3-body model with PBC. 
As shown Fig.~\ref{Fig2}, $\tilde{I}_3$ 
starts to decrease with oscillation \cite{oscillation}, and it stays negative 
indicating the spread of information over the entire system. 
We find the increase of $-\tilde{I}_3$ is logarithmic-like. 
On the other hand, figure \ref{Fig3} shows the early-time evolution of $\tilde{I}_3$ 
in the XXZ model with PBC. 
As shown Fig.~\ref{Fig3}, $\tilde{I}_3$
starts to decrease linearly with time. 
This linear light-cone decrease of the TMI is similar to the linear light-cone increase 
in BEE \cite{Kim2013} and also the operator entanglement entropy shown 
in Ref.\cite{Zhou2017}. 

Our numerical results indicate 
the time-evolution behavior of the TMI is changed by the presence of the interactions
described by $J_2$ and $J_3$ terms, i.e., from linear to logarithmic-like decrease. 
This change has also been observed in the time evolution of the BEE for many-body wave
functions with a fixed product initial state. 
In this sense, the TMI of the time-evolution operator exhibits similar behavior to 
the BEE at least in early-time dynamics. 

We also numerically investigated the mutual information $I(A:C)$, and the results
are shown in Appendix C. 
Similar behavior to the above TMI is observed by measuring  entanglement velocity 
(Tsunami velocity).  
But, please note that this correspondence does not necessarily imply that slow-dynamics 
of the system emerges with suppression of the negativity in $\tilde{I}_3$ for late-time
evolution. 
This issue will be discussed after looking at the numerical results of the late-time
evolution. 

\begin{figure}[t]
\begin{center} 
\includegraphics[width=8.5cm]{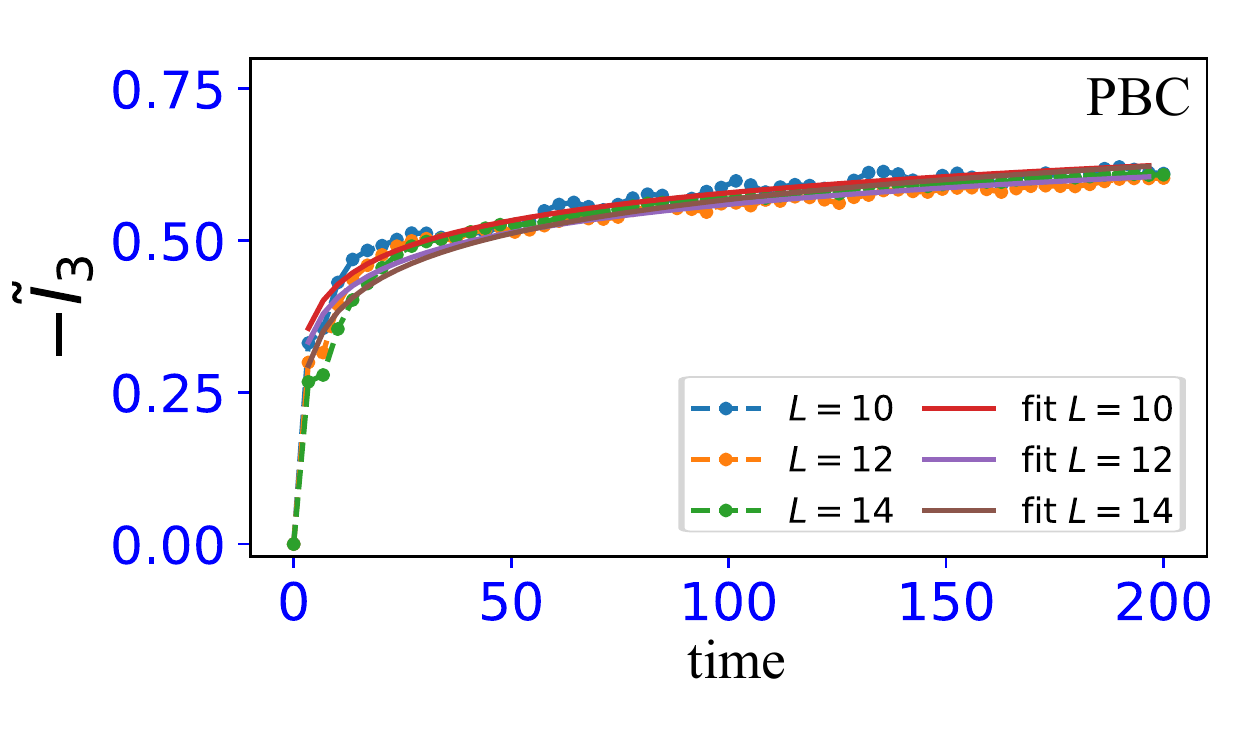}  
\end{center} 
\caption{Late-time evolution of the 3-body model. 
The fitting lines are 
$(-\tilde{I}_3)=0.0457 \log_2 t+0.2748$, $(-\tilde{I}_3)=0.0465 \log_2 t+0.2508$ and
$(-\tilde{I}_3)=0.0560 \log_2 t +0.1958$ for $L=10$, $12$ and $14$, respectively.
We set $J_1=0.3$, $J_0=2$ and $v=0.2$. The unit of time is $[2\hbar/v]$. The unit of time is $[2\hbar/v]$.
In the fitting, we used the data points within $t\in[2.5:200]$.}
\label{Fig4}
\end{figure}

\subsection{Late-time evolution of TMI}

Next, we show the numerical results of the late-time evolution of the TMI for the XXZ and 3-body models. 
Contrary to the study on large-size systems, 
our calculation includes finite-size and boundary effects, and consequently, a saturation of the TMI takes place to a certain finite value. 
Nonetheless, the detailed study of late-time evolution even for finite-size systems 
may be useful for future experiments as the system size there is obviously finite, 
and it is important to numerically observe how the target models of finite-size
systems behave specifically
compared to general expectations for infinite systems (i.e., the thermodynamic limit).

Figure \ref{Fig4} shows the late-time evolution of $\tilde{I}_3$ in the 3-body model with PBC. 
For all system sizes, the saturation of $\tilde{I}_3$ takes place with a negative value. 
We find that for long-time evolution the obtained results of $-\tilde{I}_3$ can be 
fitted by a logarithmic-like function quite well. 
Also, for this parameter regime, the system-size dependence of saturation values is small.
This implies that the saturation value of the TMI, $I_3$, in the late-time evolution
scales with $\mathcal{O}(L)$,
since the TMI of the Haar random unitary $I^{H}_3$ almost scales with $\mathcal{O}(L)$.
Thus, $\tilde{I}_3$ has only negligibly small system-size dependence.

\begin{figure}[t]
\begin{center} 
\includegraphics[width=8.5cm]{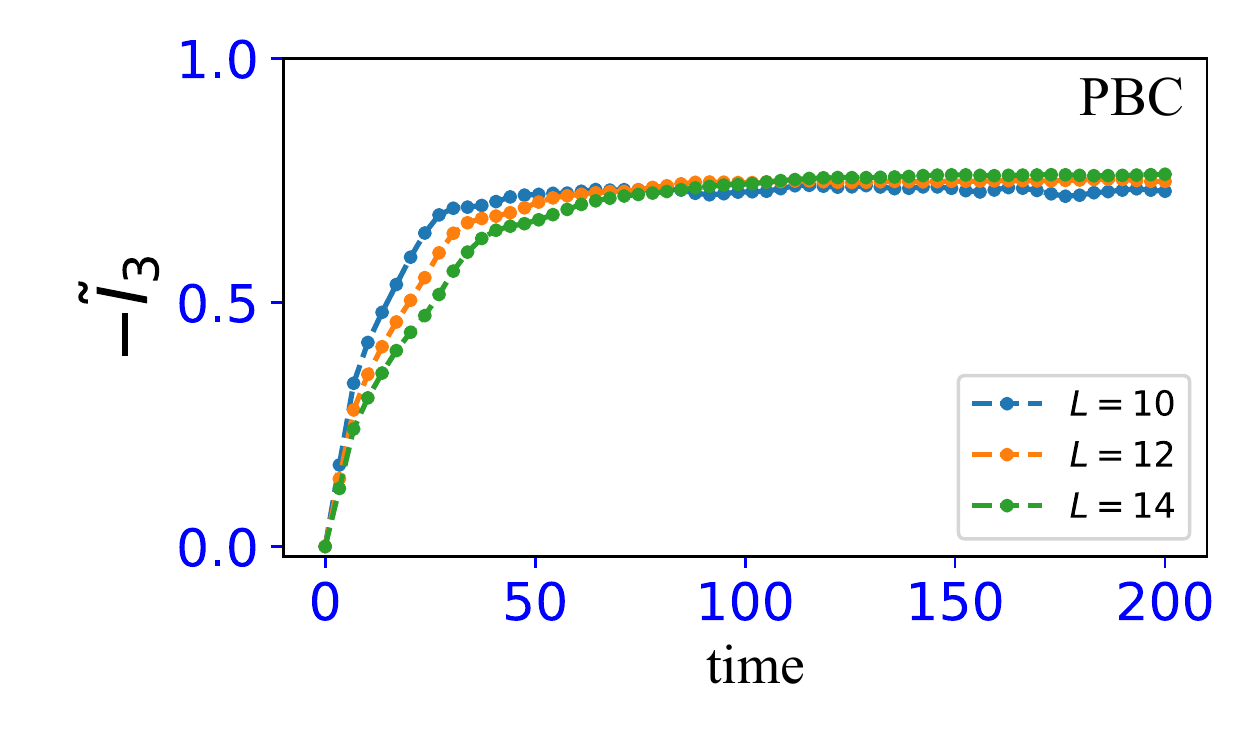}  
\end{center} 
\caption{Late-time evolution of the XXZ model
We set $v=0.2$ and $J_1=0.3$. The unit of time is $[2\hbar/v]$.
}
\label{Fig5}
\end{figure}

On the other hand, figure \ref{Fig5} shows the late-time evolution of 
$\tilde{I}_3$ in the XXZ model with PBC. 
$\tilde{I}_3$ starts to decrease linearly in early-time and it fairly slows down 
the negative growth until $(\hbar L)/2({\rm max}(v/2,J_1))$.
Finally, for both PBC and OBC cases, $\tilde{I}_3$ almost saturates with a negative 
value within the time-interval. This break-down behavior is similar to
the behavior of the OEE obtained in Ref.\cite{Zhou2017}. 

\subsection{Observation of the TMI dynamics for 4-range model}

Let us turn to the TMI in the 4-range model. 
An analytical perturbation theory in the previous work \cite{Michailidis2018} indicates
that the 4-range model can exhibit a slower-increase of the BEE compared to that 
of the 3-body model. 
We carried out the LSA with the result in Fig.\ref{LSA} (b), which shows that for large $t_0$,
the LSA deviates from the Wigner-Dyson distribution and 
it gets behavior close to that of integrable models for large $L$.

\begin{figure}[t]
\begin{center} 
\includegraphics[width=8.5cm]{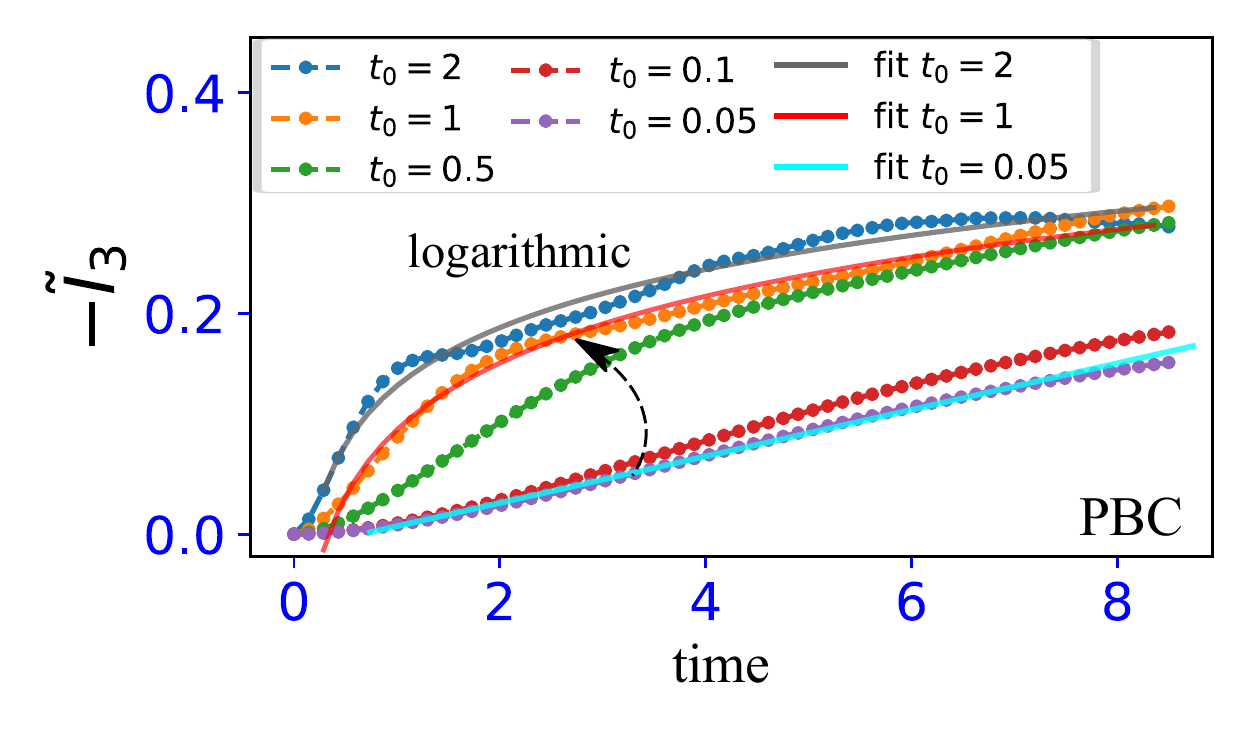}  
\end{center} 
\caption{Early-time evolution of the 4-range model. 
The fitting lines are 
$(-\tilde{I}_3)=0.0494 \log_2 t+0.2246$, 
$(-\tilde{I}_3)=0.0829 \log_2 t+0.1737$ and
$(-\tilde{I}_3)=0.0402 t-0.0260$
for $t_0=2$, $t_0=1$ and $t_0=0.05$.
We used the data points within $t\in[0.25:8.5]$ and $t\in[0.7:5.7]$ for the logarithmic
fitting and linear fitting, respectively.
For the data $t_0=2$, at the time scale $t\sim1/t_0=0.5$, 
the curvature of the behavior of the TMI takes a peak. $L=12$.
The unit of time is $[2\hbar/v]$.
}
\label{Fig6}
\end{figure}

We observe early-time evolution as varying $t_0$ with $v=0.2$. 
The results for PBC are shown in Fig.~\ref{Fig6}. 
We find that the negative growth of $\tilde{I}_3$ changes from linear-like to logarithmic-like 
as increasing $t_0$. 
In addition, we observe that for sufficiently large $t_0$ (i.e, see $t_0=2$ case), 
the negative growth tends to deviate from the logarithmic behavior.
Anyway, the multiple-spin interactions clearly affect the time evolution of the TMI in the early-time period. 
For late-time evolution for a large $t_0$, 
the decrease of the TMI is the same as that in the 3-body case, that is, the growth can be fitted by a logarithmic-like function satisfactorily (not shown). 

In addition, we also calculated the TMI under OBC with the same setup in Figs.~\ref{Fig2}, ~\ref{Fig3}, ~\ref{Fig4}, ~\ref{Fig5}, and ~\ref{Fig6}. 
The additional results are shown in Appendix C. 
These results are almost similar to those of the PBC cases.

\section{Spatial profile of spreading of TMI}
\begin{figure}[t]
\begin{center} 
\includegraphics[width=9cm]{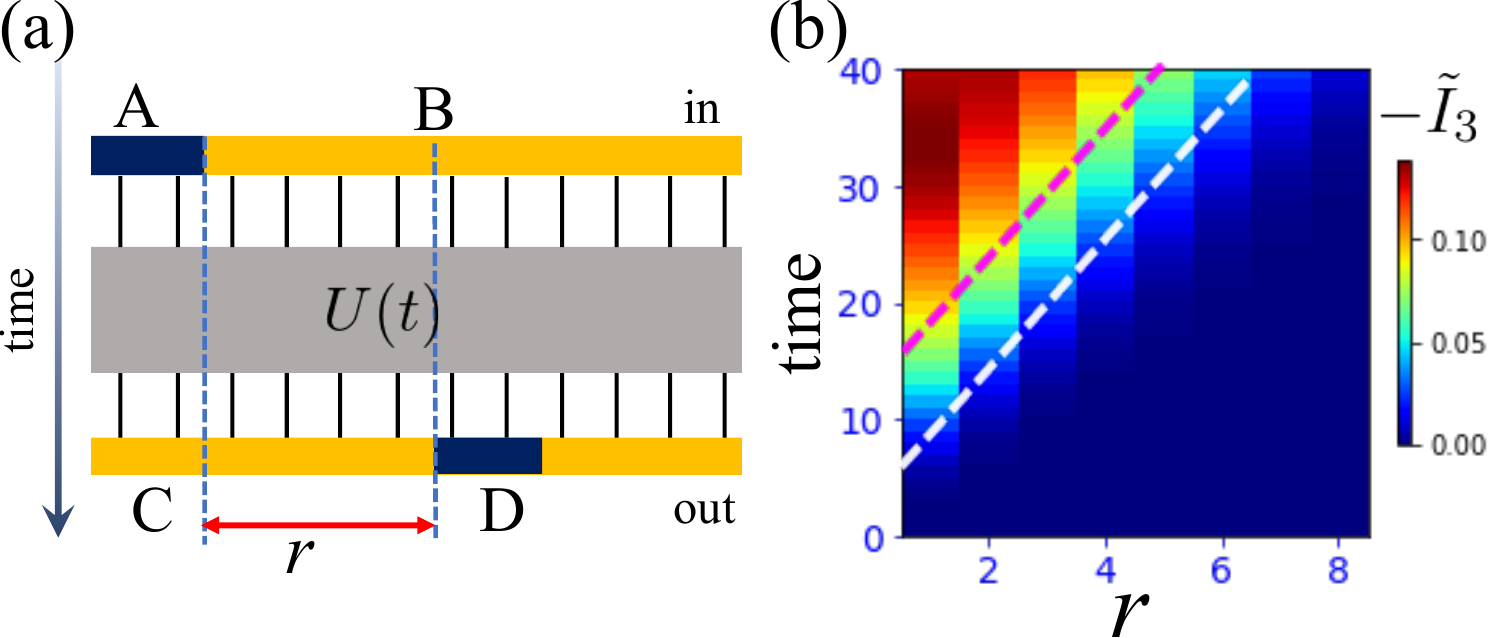}  
\end{center} 
\caption{(a) Schematic image of the partitioning of A and D subsystem with distance $r$. The subsystem A and D includes two sites. We impose open boundary condition. 
(b) Spreading of information in the XXZ model from the subsystem A to the subsystem D with distance $r$ along the time evolution. The plotted value is the TMI $-\tilde{I}_3$. For the result, we set $L=12$ and $1\leq r\leq 8$. The dashed lines are guides to the eye for approximated level lines. 
The unit of time in the vertical axis is $[2\hbar/v]$.}
\label{Fig6_2}
\end{figure}

We further investigate the spatial properties of the TMI for the three models. 
To this end, we change the partitioning of the four subsystems as shown in Fig.~\ref{Fig6_2} (a).
That is, we set both A and D to two-site subsystems, 
and study the TMI by varying the distance between them with OBC.
This setup gives qualitative insights into how the subsystems A and D separated 
with distance $r$
are correlated with each other, and how quantum information located in the subsystem A
spreads and reaches the subsystem D during the time evolution of the system.

Let us first observe $r$ dependence of the time evolution of the TMI for the XXZ model. 
Here, we calculate the TMI as a function of time for the time interval $0\leq t\leq 40$
and fix the system size as $L=12$ for various $r$'s.
The heat map result in Fig.~\ref{Fig6_2}(b) displays the spreading of 
the quantum information. 
The result indicates that quantum information seems to propagate linearly in time, consistent with the linear spreading of the TMI as shown in Fig.~\ref{Fig3}.
Therefore, even for the integrable XXZ model, the spatial 
propagation of the TMI has a linear-light cone at least in early-time dynamics.

\begin{figure}[t]
\begin{center} 
\includegraphics[width=9cm]{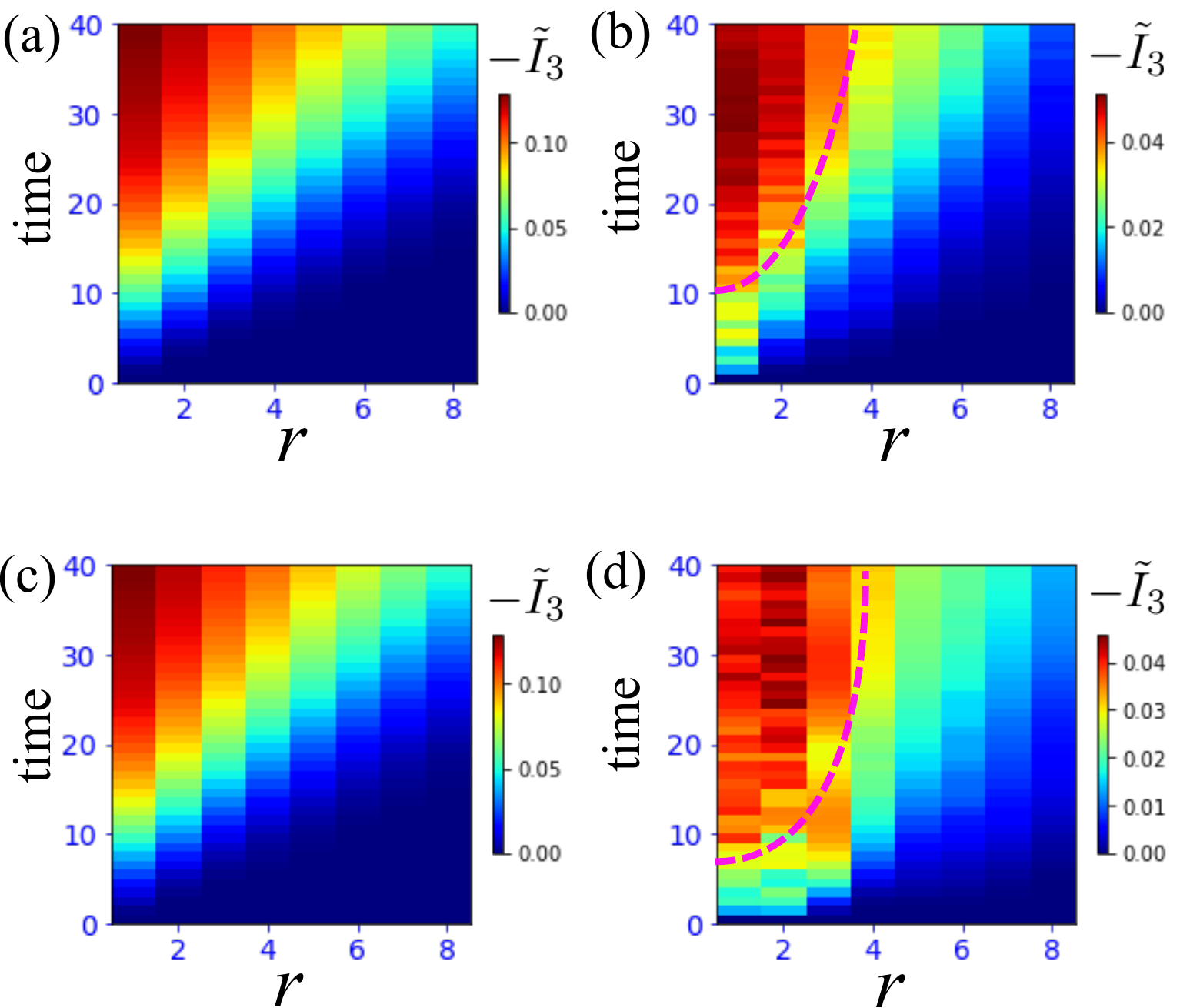}  
\end{center} 
\caption{Spreading of Information from the subsystem A to the subsystem D with distance $r$ along the time evolution of the 3-body spin model [(a) and (b)] 
and 4-range model [(c) and (d)]. 
For the data (a) and (b), chaotic case ($J_0=0.5$) and nearly integrable case ($J_0=2$), respectively. 
For the data (c) and (d), chaotic case ($t_0=0.5$) and nearly integrable case ($t_0=2$), respectively.
The plotted value is the TMI $-\tilde{I}_3$. For all results, we set $L=12$ and $1\leq r\leq 8$.
The dashed lines are guides to the eye for approximated level curves. The unit of time in the vertical axis is $[2\hbar/v]$.}
\label{Fig6_3}
\end{figure}

We next observe $r$ dependence of the time evolution of the TMI for the 3-body and 4-range models with the same setup as the case of the XXZ model. 
The heat map result of the 3-body case is shown in Figs.~\ref{Fig6_3} (a) and (b). 
The result indicates that the spatial spreading behavior of the TMI changes from linear-like to non-linear in time. 
For small $J_0$, the spreading is linear-like, while for large $J_0$, the linear-like
spreading breaks down and the TMI, $-\tilde{I}_3$, takes smaller values compared with
those of the small $J_0$ case in the same time interval. 
This behavior of the TMI is related to integrability examined by 
the LSA in Fig.~\ref{LSA} (a), that is, the non-integrability and the linear-like
spreading of the TMI correlate with each other, and similarly,
the integrable nature induced by the 2 and 3-body interactions is related
to the non-linear spreading of the TMI. 
The same observation also appears in the 4-range model as shown in Figs.~\ref{Fig6_3}(c)
and (d).
In particular, for large $t_0$, the linear-like spreading seems to break down more 
strongly than in the 3-body case. 
In addition, the spatial propagation profile of the TMI seems to relate to the ETH character obtained by the numerical results of the histogram of the local magnetization in Fig.~\ref{H_LSz}.

To summarize the results in Figs.~\ref{Fig6_2} and \ref{Fig6_3}, 
the spatial profiles of the spreading of the TMI on the setup in Fig.~\ref{Fig6_2}(a)
show the linear-like quantum scrambling in the non-integrable parameter regime
observed by the LSA, and also the non-linear propagation for the integrable 
parameter regime with large multiple spin interactions.

\section{Estimation of the late time value of TMI}
\begin{figure}[t]
\begin{center} 
\includegraphics[width=8.5cm]{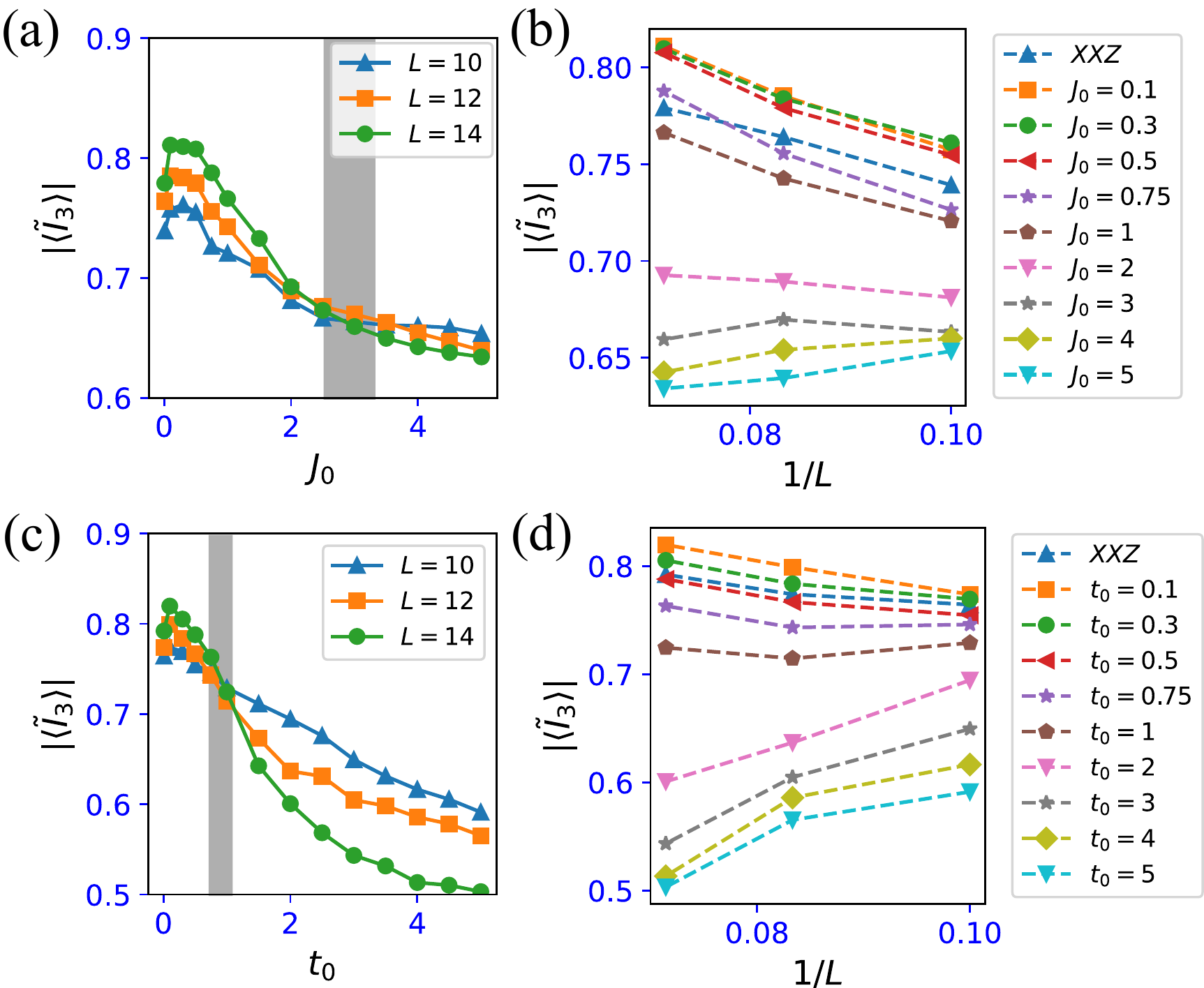}  
\end{center} 
\caption{
(a) The TMI at $t=10^5$ vs $J_0$ for the 3-body case. 
The data lines of three different system sizes seem to cross at $J_0\sim 2.5$. 
(b) The TMI vs system size $L$ for the 3-body case for each $J_0$.
We set $v=0.2$, $J_1=0.3$. 
(c) The TMI at $t=10^5$ vs $t_0$ for the 4-range case. 
In data (c), the data lines of three different system sizes cross at $t_0\sim 1$. 
(d) The TMI vs system size $L$ for the 4-body case for each $t_0$.
We set $v=0.2$.
}
\label{Fig7}
\end{figure}

In the finite-size system, it is important to see
how saturation values of the TMI in the time evolution depend on the parameters, 
system size, and integrability. 
The numerical results of the late-time evolution in the previous section exhibit almost saturating behavior of $\tilde{I}_3$. 
Here we focus on the PBC case and the value of $\tilde{I}_3$ at a very 
late time $t=10^3$, when $\tilde{I}_3$ is saturating for small $J_0$($t_0$) 
and is almost saturating for large $J_0$ ($t_0$) \cite{Nearly_sat} in 
the 3-body and 4-range models with various system sizes.
We denote the absolute value of the TMI at $t=10^3$ by $|\langle \tilde{I}_3\rangle|$. 
In the calculation, we widely vary the parameters $J_0$ and $t_0$ 
and consider the three system sizes, $L=10$, $12$, and $14$.

We summarize the calculations of $|\langle \tilde{I}_3\rangle|$ for the 3-body model in Figs.~\ref{Fig7}(a) and ~\ref{Fig7}(b) and, 
for the 4-range models in Figs.~\ref{Fig7}(c) and ~\ref{Fig7}(d), respectively.

Let us see Fig.~\ref{Fig7}(a). 
We display the numerical results of $|\langle \tilde{I}_3\rangle|$ as a function of $J_0$ 
for the three different system sizes, and find that the three curves of 
$|\langle \tilde{I}_3\rangle|$ cross with each other at $J_0\sim 2.5$. 
This indicates the existence of a phase transition from strong scrambling to weak 
or slow scrambling as $J_0$ is increased. 
Therefore, the estimation of the late time value of the TMI may be useful for 
detecting a phase transition from the viewpoint of the information spread and
scrambling, as it was already observed by studies on the measurement induced phase 
transition \cite{Zabalo2020}.
The behavior of $|\langle \tilde{I}_3\rangle|$ in Fig.~\ref{Fig7}(a) is also related 
to the integrability of the system, which is examined by the LSA: 
In Figs.~\ref{Fig7}(a) and ~\ref{Fig7}(b) for larger $J_0$, 
where the model tends to be integrable, $|\langle \tilde{I}_3\rangle|$ is suppressed. 
We furthermore note that the behavior of
$|\langle \tilde{I}_3\rangle|$ in Fig.~\ref{Fig7}(a) seems to be 
related to the ETH properties of the system in Fig.~\ref{H_LSz}(a) and \ref{H_LSz}(b).

We also plot the $|\langle \tilde{I}_3\rangle|$ vs. $1/L$ 
for each value of $J_0$ as in Fig.~\ref{Fig7}(b). 
We find that the gap of $|\langle \tilde{I}_3\rangle|$'s between $J_0=1$ and $J_0\sim 3$
is getting larger for larger $L$ (smaller $1/L$). 
This means that for larger $L$, the properties of scrambling (information spreading) 
change more clearly as a phase-transition-like manner under varying the strength of 
the interactions.

We also comment that compared to the XXZ model (nonlocalized integrable model), 
$|\langle \tilde{I}_3\rangle|$ in Fig.~\ref{Fig7}(b)
for small $J_0 \leq 0.5$ is larger than that of the XXZ model, indicating that the weak
multiple-spin interactions enhance the negativity of the TMI. 
This behavior of the TMI is consistent with the results of the LSA 
shown in Fig.~\ref{LSA}, and it implies that 
the 3-body model with weak multiple-spin interactions
exhibits non-integrable nature and the ETH tendency.
A similar phenomenon was observed in Ref.~\cite{Brenes},
which shows that the XXZ model acquires the ETH behavior by a local 
integrability-breaking perturbation, 
where Wigner-Dyson distribution of energy levels was also observed~\cite{Santos1,Santos2}.
The above trends of the TMI also appear in the case $L=12$ and $14$ cases as shown 
in Fig.~\ref{Fig7} (b).
We note that even for large $J_0$, the value of $|\langle \tilde{I}_3\rangle|$ is not strongly suppressed (fairly deviates from zero). 
This result indicates the limits on determining how strongly the TMI is suppressed by the interactions.

The above properties of the 3-body model also exist in the 4-range model 
as shown in Fig.~\ref{Fig7}(c) and ~\ref{Fig7}(d). 
We plot the $|\langle \tilde{I}_3\rangle|$ as a function of $t_0$ for the
three different system sizes in Fig.~\ref{Fig7}(c).
The three curves with different $L$'s tend to cross with each other at $t_0\sim 1$.
This again indicates the existence of a phase transition. 
However, we must be careful and do not conclude that there is a distinct phase transition 
through this finite system-size investigation accessible in our numerical resource.
Anyway,
the behavior in Fig.~\ref{Fig7} (c) also follows the results of the LSA shown in Fig.~\ref{LSA} and seems to be in good agreement with the results of the ETH properties 
shown in Figs.~\ref{H_LSz}(c) and ~\ref{H_LSz}(d).

We comment that for large $t_0$ ($\gtrsim 1$), 
the values of $|\langle \tilde{I}_3\rangle|$ are smaller than those of the 3-body 
system ($J_0$ ($\gtrsim 1$)). 
This indicates that the 4-range terms induce stronger suppression of the scrambling than 
the 3-body terms.

We furthermore plot the $|\langle \tilde{I}_3\rangle|$ vs. $1/L$ for each $t_0$ 
in Fig.~\ref{Fig7}(d), and observe the same behavior as in Fig.~\ref{Fig7}(b). 
That is,
for larger $L$, the properties of the scrambling (information spreading) change clearly 
as a phase-transition-like manner as the strength of the interaction $t_0$ is increased.

Summarizing the above observations, we conclude that the degree of integrability and ETH, measured by the LSA and the local magnetization are related to the degree of the scrambling  observed by the TMI for the present disorder-free 3-body and 4-range models. 
The results obtained in this paper do not agree with the observation in \cite{Iyoda2018}, where the non-existence of relationship between integrability and the degree of the
scrambling, but agree with the observation in 
the study in the MBL system \cite{Mascot2020,Bolter2021}, 
where the relationship between integrability and the degree of the scrambling exists.

As we emphasized in introduction, the TMI is  an efficient reliable measure for investigation
of quantum information spreading and scrambling.
We think that we obtained a clear understanding of the relationship between the integrability and information scrambling for the spin systems having parameter 
regimes close to both the ETH and MBL in this work.

Here, we comment that the multiple-spin operators
in the 3-body and 4-range models are also conserved quantities for $v=0$. 
The system with a finite $v$ approaches a nearly integrable system as these interactions are getting strong. 
However, these interactions are not single-site operators but non-local ones. 
Such a non-local conserved quantity can enhance the spread of information. 
We speculate that the combination of non-locality of these non-local multiple-spin interactions and the small hopping $H_{\rm hop}$ may inhibit strong suppression 
($|\langle \tilde{I}_3\rangle| \ll 1$). 
For this point, the previous study \cite{Michailidis2018} gave a similar expectation: 
The 3-body model indeed exhibits quasi-localization, even for large multiple-spin
interactions, and the final value of the BEE exhibits thermal value in the
thermodynamic limit.

Another comment concerns the finite system size.
One may wonder that the above observation for the existence of 
the crossover/ phase transition between the ETH and nearly-integrable regimes
is simply a finite-size effect and it disappears in the thermodynamic limit.
Very recently, however, an interesting observation was proposed in which such a kind of 
crossovers survive in a \textit{dynamic limit}~\cite{Bulchandani}.
To approach the dynamic limit, the parameters of the Hamiltonian must be scaled
with the system size.
By doing an appropriate scaling, chaos (ETH)-localization transition is to be
observed.
For the present models, the scaling law of the parameters is a future problem,
although similar quantum spin models are studied in Ref.~\cite{Bulchandani}.
However, the data of the 3-body and 4-range models shown in Figs.~\ref{Fig7}(a) 
and \ref{Fig7}(c), 
$|\langle \tilde{I}_3\rangle|$ vs $t_0$, seem to indicate a phase transition. 
That is, by suitable scaling of $t_0$ with system size $L$, the three curves may 
collapse to a single curve~\cite{Bulchandani}, although investigation for various 
system sizes is needed to verify that.\\

\section{Discussion and conclusion}

We numerically studied the spread of information and scrambling by the unitary time
evolution operator of some disorder-free spin models with multiple-spin interactions
by investigating the TMI. 
The TMI is a suitable quantitative observable to quantify the spread of information
and scrambling, and we acquired reliable observations concerning the relationship
between the quantum information scrambling, the integrability and ETH.

The target models are defined by adding integrability-breaking terms to the XXZ
model, which is a non-localized integrable model.
The additional terms describe multi-spin interactions, and by increasing the coefficients 
of these terms, the models are expected to approach the regime with 
integrable again and localized nature. 
In order to elucidate the properties of integrability and ETH for the target models, we first performed the LSA and investigated the local magnetization by numerical methods.
Both numerical results support the above expectation, that is, 
the models first deviate from the integrable regime of the XXZ model keeping scrambling nature and then move to the `phase' with properties of local integrability by the strong interactions, leading to a slow scrambling.
These two regimes have distinct properties.

Then, we moved on the study of the time evolution of the TMI.
Even in a finite-size system, the TMI exhibits non-trivial growth in early-time. 
For weak hopping and relatively strong interactions, we found the negative increase of 
the TMI is logarithmic-like 
in contrast to the linear light-cone growth in the integrable XXZ model.
The multiple-spin interactions induce logarithmic-like modification for 
the spread of information.
To verify the above observation, we furthermore studied the spatial 
evolution of the TMI by using some specific partitioning of the subsystems
for the TMI, and found that in the weak integrability-breaking regime corresponding
to the ETH, the scrambling takes place in the linear light cone in time,
whereas for the strong multi-spin interactions, this linear spreading of the TMI
breaks down.

We also numerically investigated the late-time evolution of the TMI in the finite-size
systems, where our accessible system size is up to $L=14$. 
We observed that the nearly saturation values of the TMI for various parameters in the present
disorder-free spin models depend on properties of integrability and ETH estimated by the LSA and the local magnetization. 
This implies that integrability and ETH nature of the system exhibits a strong correlation
with the degree of the scrambling measured by the TMI.
In particular, a phase-transition-like behavior of the models deduced by the
local magnetization and spatial spreading of quantum information in the early-time
evolution also emerges in the saturation value of the TMI.

A comment is in order; the non-local quasi-conserved quantity corresponding to the
multiple-spin operators in the multiple-spin interactions may inhibit strong suppression
of the TMI ($|\langle \tilde{I}_3\rangle| \ll 1$). 
The effect of the magnitude of ``$\ell$-bit" on the TMI was studied by 
using other models, and results were reported in our recent paper~\cite{Orito2022}.
Compared to the MBL case \cite{Mascot2020,Bolter2021}, 
our numerical results indicated that the strong suppression of the TMI does not occur in our models.

Nevertheless, we must be careful to take our conclusions obtained from numerical observations to be very general and definitive, because our numerical results are far from the thermodynamic limit. 
The study of the TMI for larger system sizes by alternative numerical schemes 
will be future work, in particular, to verify our findings in the study of 
late-time dynamics. 
An MPO approach \cite{Luitz2017} may be efficient for this end.\\

\section*{Acknowledgements}
This work is supported by JSPS KAKEN-HI Grant Number JP21K13849 (Y.K.) and 
T.O. has been supported by the Program for Developing and Supporting the Next-Generation of Innovative Researchers at Hiroshima University.

\bigskip
\begin{figure}[b]
\begin{center} 
\includegraphics[width=7.5cm]{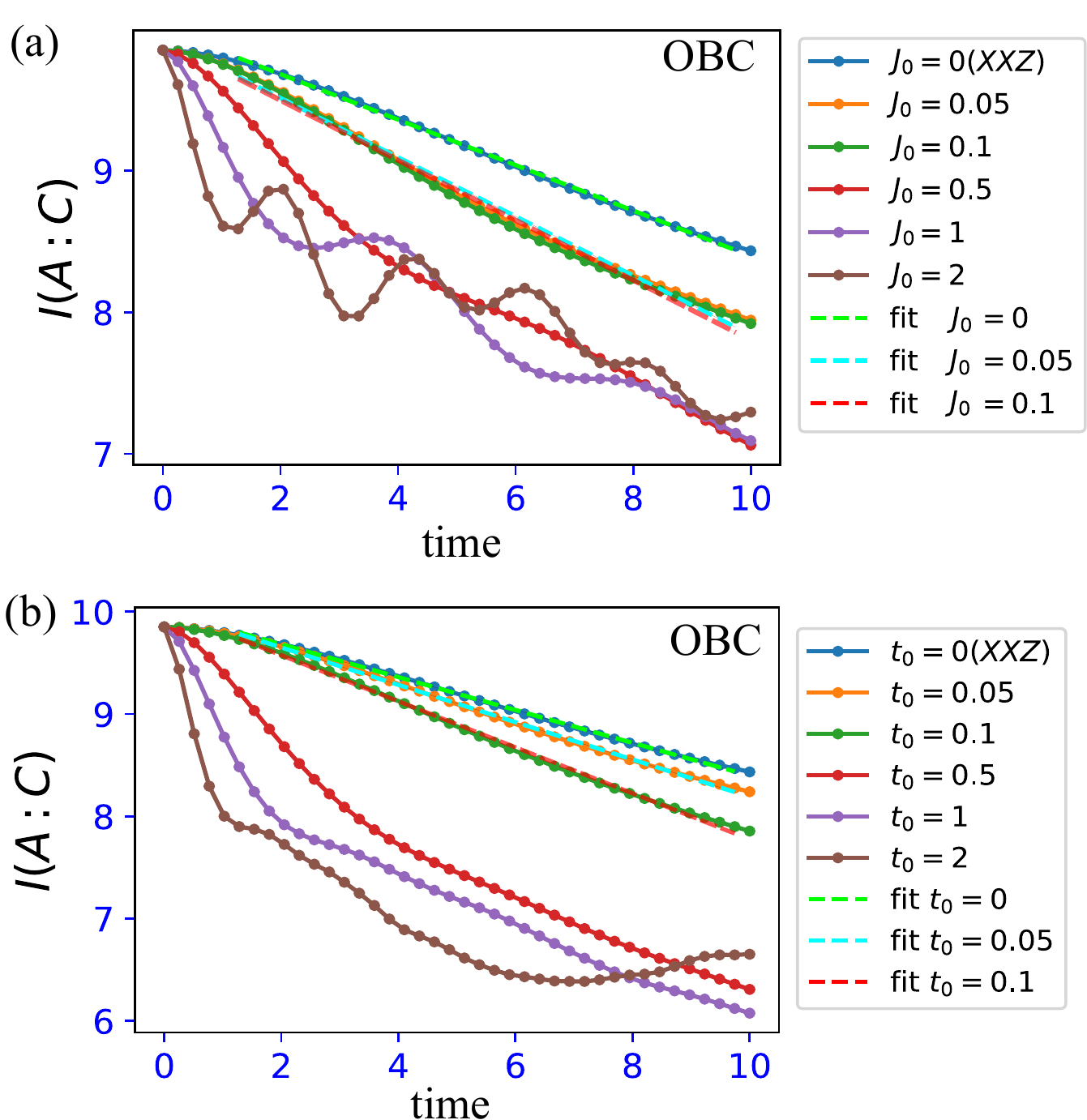}  
\end{center} 
\caption{(a) Early-time evolution of mutual information $I(A:C)$ of 3-body [(a)] and 4-range models [(b)]. We used OBC. 
In the result (a), the linear fitting lines are $I(A:C)= -0.1606 t+10.0032$, $I(A:C)=-0.2090 t +9.9318$ and $I(A:C)= -0.2118 t +9.9214$ for $J_{0}=0(XXZ)$, $J_{0}=0.05$ and $J_{0}=0.1$, respectively. 
In the result (b), the linear fitting lines are $I(A:C)=-0.1825 t+10.0157$ and $I(A:C)= -0.2250 t+ 10.0247$ for $t_{0}=0.05$ and $t_{0}=0.1$. The unit of time is $[2\hbar/v]$.}
\label{Ent_v}
\end{figure}

\appendix
\begin{figure*}[t]
\begin{center} 
\includegraphics[width=16cm]{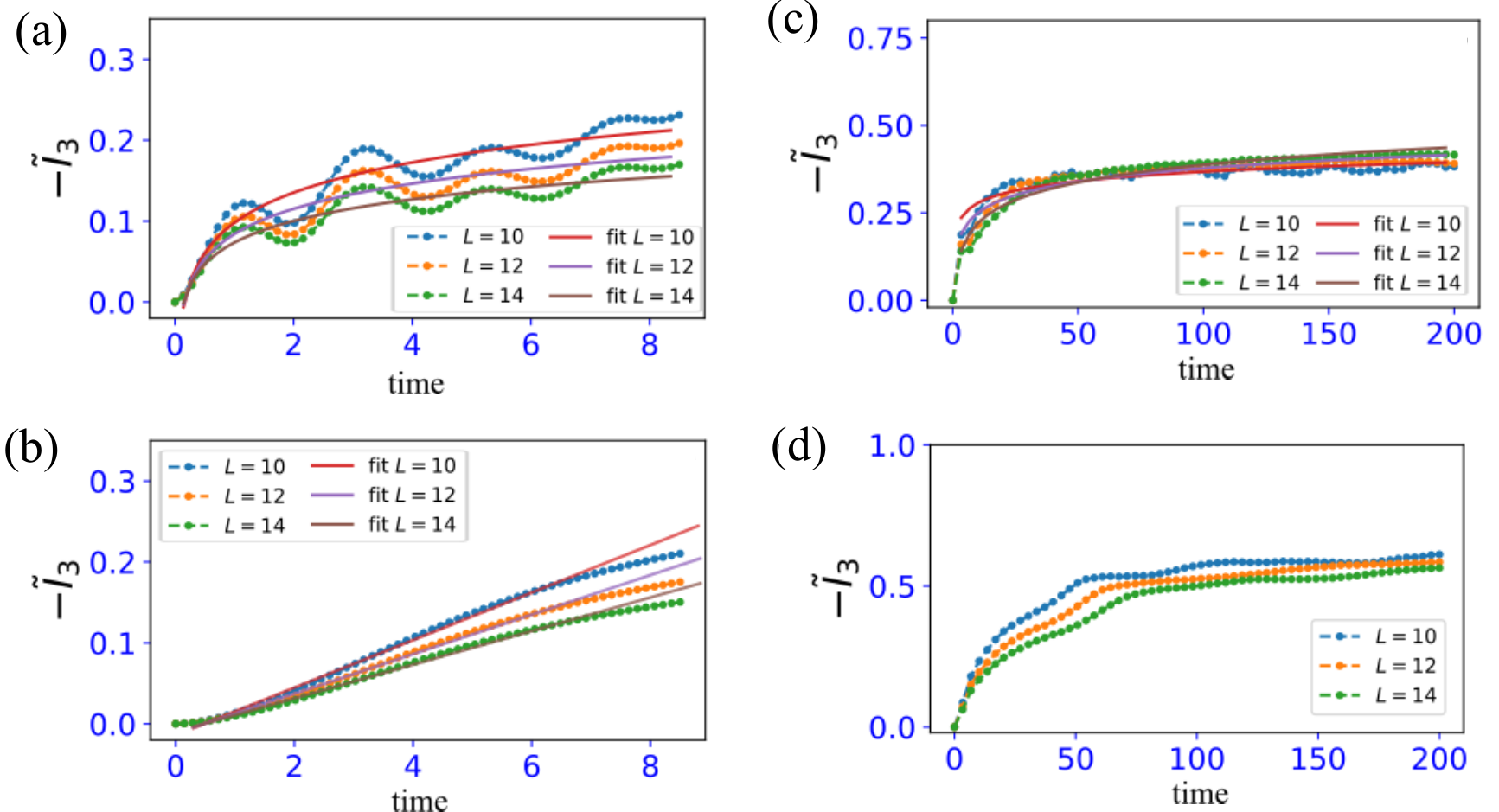}  
\end{center} 
\caption{
Additional data in open boundary condition; 
(a) Early-time evolution of 3-body model $H_{\rm 3B}$. 
The logarithmic fitting lines are 
$(-\tilde{I}_3)=0.0373 \log_2 t+0.0970$, $(-\tilde{I}_3)=0.0311 \log_2 t+0.0840$ and
$(-\tilde{I}_3)=0.0267 \log_2 t +0.0735$ for $L=10$, $12$ and $14$, respectively. We set $J_1=0.3$, $J_0=2$ and $v=0.2$. In the fitting, we used the data points within $t\in[0.1:8]$.
(b) 
Early-time evolution of XXZ model. 
The linear fitting lines are 
$(-\tilde{I}_3)=0.0295 t-0.0145$, 
$(-\tilde{I}_3)=0.0246 t-0.0121$ and
$(-\tilde{I}_3)=0.0211 t-0.0105$ for $L=10$, $12$ and $14$, respectively.
We set $v=0.2$ and $J_1=0.3$.
In the fitting, we used the data points within $t\in[0.28:6.94]$.
(c) 
Late-time evolution of the 3-body model. 
The fitting lines are 
$(-\tilde{I}_3)=0.0269 \log_2 t+0.1889$, $(-\tilde{I}_3)=0.0384 \log_2 t+0.1226$ and
$(-\tilde{I}_3)=0.0504 \log_2 t +0.0535$ for $L=10$, $12$ and $14$, respectively.
We set $J_1=0.3$, $J_0=2$ and $v=0.2$.
In the fitting, we used the data points within $t\in[2.5:200]$.
(d) Late-time evolution of the XXZ model
We set $v=0.2$ and $J_1=0.3$. The unit of time is $[2\hbar/v]$.}
\label{Fig_AppB}
\end{figure*}
\section*{Appendix A: Time independent part of entropy in calculation of the TMI}
In this work, we focus on zero-magnetization sector of the spin Hilbert space, 
with not $2^{L}$ dimension but $\binom{L}{L/2}$-dimension and consider that 
all subsystems A, B, C, and D are equal, that is, including $L/2$-lattice sites. 
Then, the OEE $S_{X_1}$ ($X_1=A,B,C,D$) is given by 
\begin{eqnarray}
S_{X_1}=-\sum^{L/2}_{n_D=0}
\binom{L/2}{n_D}g(n_{D})\log_2 g(n_D),
\label{SX}
\end{eqnarray}
where $g(n_D)= \binom{L/2}{n_D}/N_{D}$ ($N_D$ is total Hilbert space dimension, $\binom{L}{L/2}$).

As far as all subsystems A, B, C and D are equal, the $S_{X}$ is time-independent \cite{Hosur,Bolter2021}, hence, we only need to calculate $S_{AC}$ and $S_{AD}$, 
which are time-dependent in the calculation of the TMI.

\section*{Appendix B: Entanglement velocity in 3-body and 4-range models} 

In this appendix, we numerically observe the time evolution of the mutual information
$I(A:C)$ defined by Eq.(2) in Sec.~II.
In dynamics in general chaotic system, $I(A:C)$ starts from a certain finite value
\cite{Hosur} and then linearly decreases. 
Therefore, $I(A:C)$ behaves as $I(A:C)(t)=I(A:C)(t=0)-v_E s t$, where $s=2$ 
in spin $1/2$ bases 
and $v_E$ is entanglement velocity (sometimes called Tsunami velocity).

We investigate whether or not such a linear decrease appears in our model or how such a
linear decrease changes by varying the parameters $J_0$ and $t_0$ in the 3-body or 4-range models. 
We here focus on early-time evolution under OBC, and focus on $L=12$ system size.

Figure~\ref{Ent_v} is the time evolution of $I(A:C)$ for various parameters. 
For all data in both Fig.~\ref{Ent_v} (a) and Fig.~\ref{Ent_v} (b), $I(A:C)$ starts 
from a constant value obtained by Eq.~(\ref{SX}) and $S_{AC}=0$ at $t=0$. 
See the 3-body case in Fig.~\ref{Ent_v} (a), for small $J_0$, $I(A:C)$ almost linearly decreases and a linear fitting can be applied and the entanglement velocity can be extracted. The value $v_E$ obtained by the data is close to the hopping value $v/2$. 
But for large $J_0$, the linear decrease breaks down and the linear fitting is of course no longer applied or we cannot extract $v_E$, non-trivial decrease with oscillation appears. 
We expect that this behavior comes from the non-locality of the multiple-spin interactions, i.e., not only NN `hopping of domain walls' but also multi-distance terms
give non-trivial effects on the short-range scrambling of the information.

The same behavior is observed in the 4-range model as shown in Fig.~\ref{Ent_v} (b). 
For small $t_0$, the linear fitting of $I(A:C)$ can be applied and the entanglement velocity can be extracted. 
The value $v_E$ obtained by the data is close to the hopping value $v/2$. But for large $t_0$, such a linear decrease breaks down.

\section*{Appendix C: Additional data of TMI}
In this Appendix, we show additional data of the TMI of the XXZ and 3-body models under open boundary condition (OBC). The setup is same to the PBC case in the main manuscript. 
The results in OBC corresponding to the results in Figs.~\ref{Fig2}, ~\ref{Fig3}, ~\ref{Fig4} and ~\ref{Fig5} are shown in Figs.~\ref{Fig_AppB}(a)-(d). 

\begin{figure}[t]
\begin{center} 
\includegraphics[width=8cm]{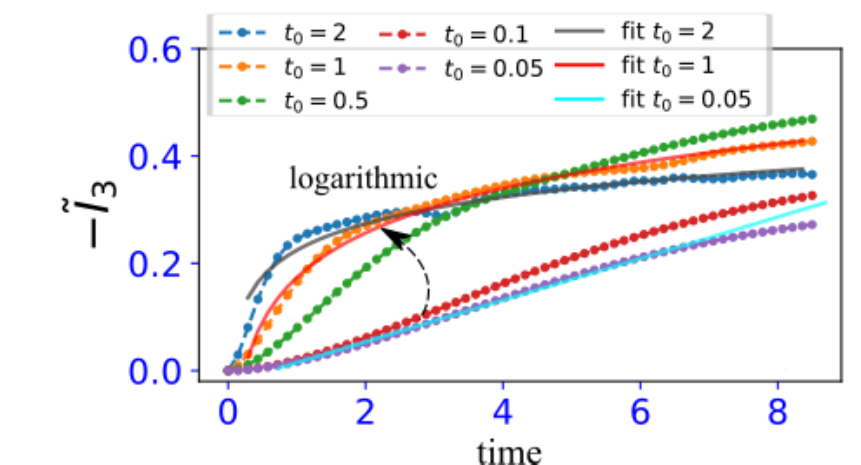}  
\end{center} 
\caption{
Early-time evolution of the 4-range model with OBC.
The fitting lines are 
$(-\tilde{I}_3)=0.0527 \log_2 t+0.1344$, 
$(-\tilde{I}_3)=0.0606 \log_2 t+0.0945$ and
$(-\tilde{I}_3)=0.0217 t-0.0158$ 
for $t_0=2$, $t_0=1$ and $t_0=0.05$.
We used the data points within $t\in[0.25:8.5]$ and $t\in[0.7:5.7]$ for the logarithmic
fitting and linear fitting, respectively.
For the data $t_0=2$, at the time scale $t\sim1/t_0=0.5$, 
the curvature of the behavior of the TMI takes a peak. $L=12$.
}
\label{Fig_AppB2}
\end{figure}

The behavior of every data is almost similar to that of the PBC case.
However, we note that for the OBC case in Fig.~\ref{Fig_AppB}(d), 
the system-size dependence of $\tilde{I}_3$ 
is somewhat larger than that of the PBC case in Fig.~\ref{Fig5}.
We expect that this comes from finite-size and boundary effects.

we further show data of the TMI of the 4-range model under OBC. 
The setup is same to the PBC case in the main manuscript. 
The result in OBC corresponding to the result in Fig.~\ref{Fig6} is shown in Fig.~\ref{Fig_AppB2}. 
The behavior is also almost similar to that of the PBC case.


\begin{thebibliography}{99}
\bibitem{Haar_ND}
$I^{H}_3$ is calculated under the dimension $2^L$ including full-sectors of total magnetization.
\bibitem{oscillation}
This oscillation comes from the multiple-spin interaction terms. 
If we set $v=0$, the TMI in the 3-body and 4-range models exhibits oscillating behavior.
\bibitem{Nearly_sat}
For large $J_0$ and $t_0$, the behaviors of the value of the TMI $\tilde{I}_3$ in 
the 3-body and 4-range models are much similar to that of the conventional MBL, 
that is, logarithmic-increase in time. 
The value of the TMI $\tilde{I}_3$ can continue to decrease to a thermal value in the
infinite time limit. 
However, at $t=10^3$, the decrease of the TMI $\tilde{I}_3$ appears to be very small 
and almost saturated.
\end{thebibliography}
\end{document}